\documentclass[a4paper,fleqn,usenatbib]{mnras}
\usepackage{newtxtext,newtxmath}
\usepackage[T1]{fontenc}
\usepackage{ae,aecompl}
\usepackage{epsfig}
\usepackage{amsmath, amsfonts, epsfig, xspace}
\usepackage{natbib}
\usepackage{deluxetable}
\usepackage{rotating}
\usepackage{graphicx}
\usepackage{caption}
\usepackage{longtable}
\usepackage{multicol}
\usepackage{booktabs}
\usepackage{dcolumn}
\usepackage{longtable}

\usepackage[normalem]{ulem}
\usepackage{booktabs}
\usepackage{threeparttable}
\usepackage{subcaption}
\usepackage{changepage}

\usepackage{stfloats}
\usepackage{cuted}
\usepackage{capt-of} 
\usepackage{booktabs}
\usepackage{array}
\usepackage{tabularx}

\def\apjl{{ApJ}}%
\def\apjs{{ApJS}}%
\def\aap{{A\&A}}%
\title[INOV study of LPSQs]{Intranight optical monitoring of eight quasars with superluminal jets and low optical polarization: Probing the mechanism of rapid flux variability}
\author[$Chand$ et al.]{{ \large Krishan Chand$^{1,2}$\thanks{E-mail: krishanchand.kc007@gmail.com (KC)},
Gopal-Krishna$^{3}$, Ashutosh Tomar$^{4,5}$\thanks{Deceased. Ashutosh Tomar (1998-2025) contributed significantly to the observations presented in this paper. This paper is dedicated to his memory.}, Vibhore Negi$^{6}$, Hum Chand$^{1}$, and Suvendu Rakshit$^{4}$}\\\\
$^{1}$Department of Physics and Astronomical Science, Central University of Himachal Pradesh (CUHP), Dharamshala 176215, India\\
$^{2}$Manipal Centre for Natural Sciences, Manipal Academy of Higher Education, Manipal, India\\
$^{3}$UM-DAE Centre for Excellence in Basic Sciences, Vidyanagari, Mumbai 400098, India\\
$^{4}$Aryabhatta Research Institute of Observational Sciences (ARIES), Manora Peak, Nainital 263002, India\\
$^{5}$Department of Applied Physics/Physics, Mahatma Jyotiba Phule Rohilkhand University, Bareilly 243006, India\\
$^{6}$Kavli Institute for Astronomy and Astrophysics, Peking University, Beijing, 100871, People's Republic of China\\
}

\begin{document}
\date{Accepted ---; Received ---; in original form ---}

\pagerange{\pageref{firstpage}--\pageref{lastpage}} \pubyear{2026}

\maketitle

\label{firstpage}

\begin{abstract}
Intranight optical variability (INOV) of blazars whose emission arises predominantly from a relativistic jet, 
has been widely linked to a combination of two key observational diagnostics: the fractional optical polarization ($P_{opt}$) and the Doppler boost arising from relativistic beaming of the jet's emission. {To discern} which of these two factors plays the dominant role, 
we investigate INOV of a well-defined representative sample of 8 flat-spectrum radio quasars (FSRQs{, which are `low-synchrotron-peaked' blazars}) whose jets exhibit a low degree of polarization ($P_{opt}$ < 3$\%$) but a large superluminal motion with c$\beta_{app}$ = 5c $-$ 31c (median = 10c).
We report sensitive optical monitoring of these 8 `low-polarization superluminal quasars' (LPSQs) using a 1.3-metre telescope, in 20 sessions of minimum 3-hour duration. These data were supplemented with similarly high-quality monitoring data/results {available in the literature for this sample, thus adding another 19 sessions.}
{Duty cycle (DC) of strong INOV (amplitude $\psi > 5\%$) is found to be zero, applying the normally used $F_{\eta}$ test, and $\sim 7\%$ using the more powerful `enhanced $F$-test'. The small DC,} despite the strong Doppler boost expected from the observed high superluminality, shows that strong INOV is much rarer among LPSQs, compared to high-polarization quasars, for which DC $(\psi > 5\%$) is found to be around 30\%, {even using the $F_{\eta}$ test}. This provides the first compelling evidence that optical polarization is the primary diagnostic of INOV, while relativistic beaming plays a supportive role. Possible implications for the proposed mechanisms for INOV are pointed out.

\end{abstract}
\begin{keywords}
galaxies: active - galaxies: photometry - galaxies: jets - quasars: general - galaxies: quasars: supermassive black holes - galaxies: nuclei. 
\end{keywords}

\section{Introduction}
\label{introduction}

Optical micro-variability, or intranight optical variability (INOV) is now widely accepted as a potent diagnostic of blazar type activity in several prominent classes of Active Galactic Nuclei (AGN) (\citealp{Gopal-Krishna2018BSRSL..87..281G}, and references therein; \citealp{Alok2018Galax...6....1G} and references therein; \citealp{Bhattacharya2019MNRAS.483.3382B}; \citealp{Pasierb2020MNRAS.492.1295P}; \citealp{Sapna2021MNRAS.507L..46M}; 
\citealp{Vineet2021MNRAS.501.4110O}).
 
Therefore, the physical mechanism underlying this phenomenon merits a close scrutiny. As per current understanding, at least in blazars, INOV is believed to arise mainly from the relativistic jet, and is usually linked to two key factors: (i) excitation/enhancement of turbulence within the jet's synchrotron plasma in the wake of relativistic shock(s) propagating along the jet, which simultaneously enhances the jet's radiative output and its degree of linear polarization (\citealp{Marscher2014ApJ...780...87M} and references therein; \citealp{Calafut2015JApA...36..255C}; also \citealp{Laing1980MNRAS.193..439L}; \citealp{Goyal2012AA...544A..37G}; \citealp{Raiteri2025A&ARv..33....8R}) and (ii) the bulk Doppler factor $\delta_j$ of the post-shock, turbulence-enhanced region in the jet flow. While the former condition (turbulence) appears crucial for seeding the micro-variability (optical fluctuations) within the jet, the Doppler boost arising from the jet's bulk relativistic motion can be crucial for enhancing the detectability of the intrinsic micro-variations.\\

A strong link of INOV (\citealp{GopalKrishna1993A&A...271...89G}), to fractional optical polarization ($P_{opt}$) was first clearly demonstrated by \citet{Goyal2012AA...544A..37G}, who also showed that a flat/inverted radio spectrum is by itself no guarantee of detecting a strong INOV (amplitude $\psi > 3-4\%$) (see also, \citealp{Stalin2004JApA...25....1S}; \citealp{Gopal-Krishna2018BSRSL..87..281G}). 
For this, \citet{Goyal2012AA...544A..37G} carried out a program of sensitive, high-cadence optical monitoring of 21 radio-loud quasars, including 9 high- and 12 low-polarization quasars (HPQs and LPQs), adopting the conventional dividing line at $P_{opt}$ = 3\% (e.g. \citealp{Stockman1984ApJ...279..485S}; \citealp{Krishan2022MNRAS.516L..18C}). Remarkably, the HPQ subset showed strong INOV (amplitude $\psi > 4\%$) in 11 out of 29 intranight sessions, in stark contrast to the LPQs for which strong INOV was observed on just one out of 44 nights. This clearly established high $P_{opt}$ as a key attribute of the radio quasars showing strong INOV. But, is this alone a sufficient marker for expecting strong INOV? What about the role of the afore-mentioned second factor, namely, the nuclear jet's Doppler factor, $\delta_j$, as inferred usually from the apparent motion of the VLBI radio knots (\citealp{Liodakis2018ApJ...866..137L} and references therein; \citealp{Britzen2018MNRAS.478.3199B}; \citealp{Lister2019ApJ...874...43L}) and from radio flux variability (e.g. \citealp{Liodakis2021A&A...654A.169L} and references therein; \citealp{Singal1985MNRAS.215..383S}; \citealp{Readhead1994ApJ...426...51R})? The above issue was recently examined by \citet{Vibhore2023MNRAS.524L..66N}, via an INOV campaign for a sample of 6 blazars. Like all blazars, they too are radio-loud  (e.g. \citealp{Kellermann2016ApJ...831..168K}). In addition, they belong to the HPQ class, exhibiting a {high polarization ($P_{opt} > 3\%$)} in the RoboPol survey (\citealp{Blinov2021MNRAS.501.3715B}) {and, moreover, have all been detected at TeV energies}. A distinctive feature of these `TeV blazars' is that the apparent speeds (c$\beta_{app}$) of their VLBI knots had been shown to be consistent with no motion relative to the core. For such blazars, \citet{Vibhore2023MNRAS.524L..66N} found a complete lack of INOV detection down to $2-3\%$ level. This showed that the presence of dominant superluminal radio knot(s) in the parsec-scale jet constitutes a key ingredient for INOV detection and while a high $P_{opt}$ is an important diagnostic, as demonstrated in \citet{Goyal2012AA...544A..37G}, it alone is not a sufficient condition for INOV detection. The rather tight correlation of INOV with $\beta_{app}$ (which is usually associated with shocks forming within the jet flow) also suggested that the zone of turbulence trailing behind the advancing shock,  
where the INOV is likely to originate (e.g. \citealp{Marscher2008Natur.452..966M}; \citealp{Calafut2015JApA...36..255C}), remains kinematically coupled to the shock, at least in this type of blazars.

An important question stemming from the above-mentioned findings is which of the above two observational factors has the primacy in the context of the observed INOV. Is it a high $P_{opt}$, or, alternatively, a high superluminality of the nuclear jet? For an explicit check on this, we pose here the question whether a strong INOV would be observed among the radio quasars whose parsec-scale jets do have a highly relativistic bulk motion, as evidenced by their highly superluminal radio knots, but which have also shown a very low $P_{opt}$ (always < 3\%). Detection of a strong INOV in such quasars would provide a strong evidence in favour of relativistic beaming being the primary factor behind INOV detection, instead of a high $P_{opt}$. Thus, the present study involves an INOV search among an unbiased sample of 8 bright low-polarization radio quasars ($P_{opt} < 3\%$), each of which shows a high superluminal motion, with c$\beta_{app}$ (5c $-$ 31c, median = 10c). 
Such quasars will be called `low-polarization superluminal quasars' (LPSQs). In the next section, we describe the selection of a representative sample of 8 LPSQs for which the results of intranight optical monitoring are presented in this study.

\section{Sample selection}
\label{sample}

\begin{table*}
\centering
\begin{minipage}{180mm}
\caption{Basic information on the selected representative sample of 8 low-polarization superluminal quasars (LPSQs).} \par

\label{tab:Table1}
 \resizebox{\textwidth}{!}{
  \begin{tabular}{cccccccc}
\hline\\
\multicolumn{1}{c}{Source}  & \multicolumn{1}{c}{Other name} & \multicolumn{1}{c}{Type} & \multicolumn{1}{c}{Redshift}  & \multicolumn{1}{c} {App. magnitude} & \multicolumn{1}{c}{Radio sp. index ($\alpha$)} & \multicolumn{1}{c}{Optical polarization}& \multicolumn{1}{c}{Apparent speed }  \\
\multicolumn{1}{c}{SDSS name}  & \multicolumn{1}{c}{} &  \multicolumn{1}{c}{} & \multicolumn{1}{c}{($z$)} & \multicolumn{1}{c}{(R)} & \multicolumn{1}{c}{$f_{\nu} \propto \nu^{\alpha}$} & \multicolumn{1}{c}{$P_{opt}$ (\%)}& \multicolumn{1}{c}{$\beta_{app}$}  \\
(1)  & (2) & (3)  & (4)  & (5) & (6)  & (7) & (8)  \\\\

\hline\\

  J074110.70$+$311200.23              & OI 363  & FSRQ$^R$  &  0.631   & 16.3  & 0.29     &  0.2$\pm$0.2    & 10.70$\pm$1.20$^a$       \\
    &           &          &        &     &     &                  &                  \\
  J084124.36$+$705342.17             & 4C 71.07  & FSRQ$^R$/LSP$^{Y,S}$ &  2.172   & 16.8  & $-$0.33  &  1.1$\pm$0.5     & 21.51$\pm$0.67$^a$          \\
   &           &          &        &      &    &                  &               \\
  J115043.87$-$002354.20              & 4C$-$00.47 & FSRQ$^R$ &  1.980   & 17.1  & $-$0.48  &  0.7$\pm$0.7     &  9.20$\pm$1.90$^a$    \\
    &           &          &        &      &    &                  &                      \\
  J122906.69$+$020308.59              & 3CR 273  & FSRQ$^R$/LSP$^{Y,S}$ &  0.158   & 14.1  & $-$0.01  &  0.5$\pm$0.1     & 14.91$\pm$0.16$^a$     \\
    &           &          &        &      &    &                  &                  \\
 J135704.44$+$191907.37              & 4C 19.44  & FSRQ$^R$/LBL$^S$ &  0.720   & 15.3  & $-$0.15  &  0.3$\pm$0.3     & 9.84$\pm$0.70$^b$     \\
    &           &          &        &    &      &                  &                  \\
  J161341.06$+$341247.91              & OS 319   & FSRQ$^R$/LSP$^{Y,S}$ &  1.398   & 17.1  & 0.11     &  1.7$\pm$0.7     & 31.10$\pm$1.90$^a$    \\
    &           &          &        &     &     &                  &                  \\
  J213638.59$+$004154.21           & PHL 61  & FSRQ$^R$  &  1.941  & 17.0   & 0.77     & 0.4$\pm$0.4      & 5.10$\pm$0.30$^c$           \\
    &           &         &         &    &      &                  &                \\
  J220314.98$+$314538.27              & 4C 31.63  & FSRQ$^R$/LSP$^S$ &  0.295  & 14.3   & 0.20     &  0.2$\pm$0.1     &  8.67$\pm$0.12$^a$         \\
    &           &           &         &        &      &            &                  \\
 
\hline\\

\multicolumn{8}{l}{References: (R) ROMA-BZCAT \citep{Massaro2015ApSS.357...75M}; (Y) \citet{Yang2022ApJS..262...18Y}; (S) SSDC Multi-Catalogue Search (see Footnote 1 in Section \ref{sample}) }\\
\multicolumn{8}{l}{(a) \citet{Homan2021ApJ...923...67H}; (b) \citet{Marshall2018ApJ...856...66M}; (c) \citet{Homan2015yCat..17980134H}.}\\
  \end{tabular}}
 \end{minipage}
\end{table*}

The present set of 8 LPSQs was extracted from the well-defined sample of 41 low-polarization radio quasars (LPRQs) analysed in \citet{Krishan2022MNRAS.516L..18C}. That sample had been derived from the opto-polarimetric study by \citet{Impey1990ApJ...354..124I}, using 140 radio-loud quasars which they had selected from the 1-Jy catalogue of 518 sources defined by \citet{Kuehr1981AAS...45..367K} at 5 GHz. 
Out of these 140 quasars, \citet{Krishan2022MNRAS.516L..18C} short-listed only those for which optical polarization information was available in \citet{Impey1990ApJ...354..124I}, and ZTF light-curves (LCs) in at least one of the two bands ($g$ and $r$) were also available,
with minimum 25 data points and a noise level $\sigma$ (median) < 0.15 mag. 
This left 90 quasars, out of which 41 could be classified as LPRQs, based on the maximum fractional polarization of $<3\%$, as reported by \citet{Impey1990ApJ...354..124I} and \citet{Impey1991ApJ...375...46I} (see \citealp{Krishan2022MNRAS.516L..18C} for details). Among this set of 41 LPRQs, our choice was limited to those lying north of $-$5 degrees declination and having R-mag < 17.5, from consideration of observational feasibility. This filter led to a set of 21 LPRQs. Two of these sources, namely J075052.05$+$123104.83 and J163515.49$+$380804.50, had to be discarded as they had exhibited polarization class transitions (\citealp{Krishan2022MNRAS.516L..18C}).
Another exclusion was the source J054236.14$+$495107.23,  since the one-sigma error on its published $P_{opt}$ places it at the LPRQ/HPRQ devide, rendering its LPRQ status uncertain. This sequence of selection filters left us with a set of 18 LPRQs. Out of these, J192748.49$+$735801.57, had to be left out because its intranight monitoring for a few hours was not feasible using the telescopes available to us, as its nightly transit occurs during the monsoon time. Finally, for the remaining 17 sources, we scrutinised the literature to check availability of $\beta_{app}$ and 8 of them found to have c$\beta_{app} > 5$c were selected, after discarding two in view of their steep radio spectra. Basic information on this sample of 8 flat-spectrum LPSQs is summarized in Table \ref{tab:Table1}. {Column 1 gives the SDSS source name, while column 2 lists alternative common identifiers. Source classifications (FSRQ, LSP, or LBL) are provided in column 3, based on ROMA-BZCAT \citep{Massaro2015ApSS.357...75M}, \citet{Yang2022ApJS..262...18Y} and SSDC Multi-Catalogue Search{\footnote{\url{https://tools.ssdc.asi.it/mmc.jsp}}}. Redshifts ($z$) from NED and R-band magnitudes from SIMBAD are given in columns 4 and 5, respectively. Columns 6 and 7 list the radio spectral indices ($\alpha$) and optical polarizations ($P_{opt}$) from \citet{Krishan2022MNRAS.516L..18C}. Finally, column 8 presents the apparent speed ($\beta_{app}$) in units of c compiled from \citet{Homan2015yCat..17980134H}, \citet{Marshall2018ApJ...856...66M} and \citet{Homan2021ApJ...923...67H}.}

\section{Photometric Monitoring and Data Analysis}
Intranight photometric monitoring of 7 out of the 8 LPSQs in our sample was carried out in the Johnson-Cousins $R$-band using the 1.3-metre Devasthal Fast Optical Telescope (DFOT, \citealp{Sagar2011CSci..101.1020S}), operated by the Aryabhatta Research Institute of observational sciencES (ARIES), Nainital, {India}. The INOV data for the remaining source, namely J135704.44$+$191907.37 were abundantly available in \citet{Goyal2012AA...544A..37G} and we have used those data here. In our observations, each source was monitored in at least three sessions of minimum 3-hour duration, except for J084124.36$+$705342.17 and 122906.69$+$020308.59 (3C 273), for which only one session longer than 3-hour could be realised by us. {Nonetheless}, additional INOV data for both LPSQs could be availed from the literature, as described below. Thus, monitoring during a total of 20 intranight sessions was newly carried out by us, between January 2024 and April 2025, using DFOT. Details of the observational set-up and the procedure followed can be found in \citet{Sapna2019MNRAS.489L..42M} and \citet{Krishan2022MNRAS.511L..13C}.

\begin{figure*}

   \includegraphics[width=1.08\textwidth,height=0.95\textheight]{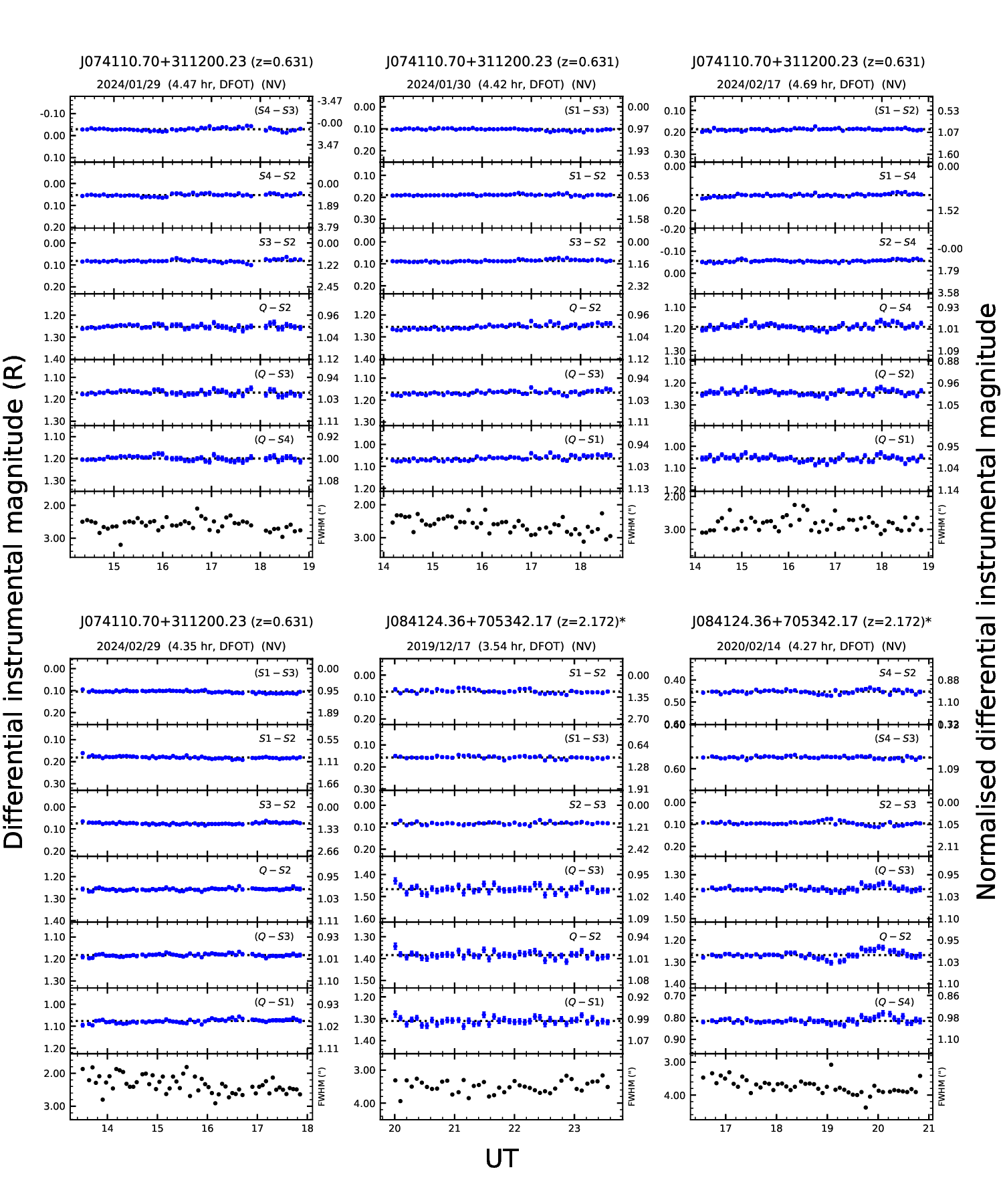}\\
   
  \vspace{-0.09in}
  \caption{Differential light curves (DLCs) for our sample of 8 low-polarization superluminal quasars (LPSQs), based on the analysis carried out in the present study. At the top of each panel, the source name, its redshift and the date $\&$ duration of the monitoring session are mentioned, together with the name of the telescope used and the INOV status determined for the session. In each mosaic, the upper three panels show the DLCs for the pairs of the comparison stars. The subsequent lower panels show the DLCs of the target LPSQ relative to the three comparison stars, as indicated by the labels {(The labels inside parentheses refer to the DLCs using the steadiest two comparison stars)}. The lowermost panel displays the variation of the seeing disk (FWHM), i.e., `point spread function' (PSF), through the monitoring session. On the right side of all panels, the differential instrumental magnitude ({R-band}) is shown, normalised by its mean value {for the session}. The dotted horizontal line marks the corresponding mean level. An asterisk, `*' in the header means that the DLCs and other data have been taken from the literature. The particulars of the comparison stars used are given in Table \ref{tab:comparison_color}.}
\label{fig1} 
\end{figure*}

\begin{figure*}

  \includegraphics[width=1.08\textwidth,height=0.95\textheight]{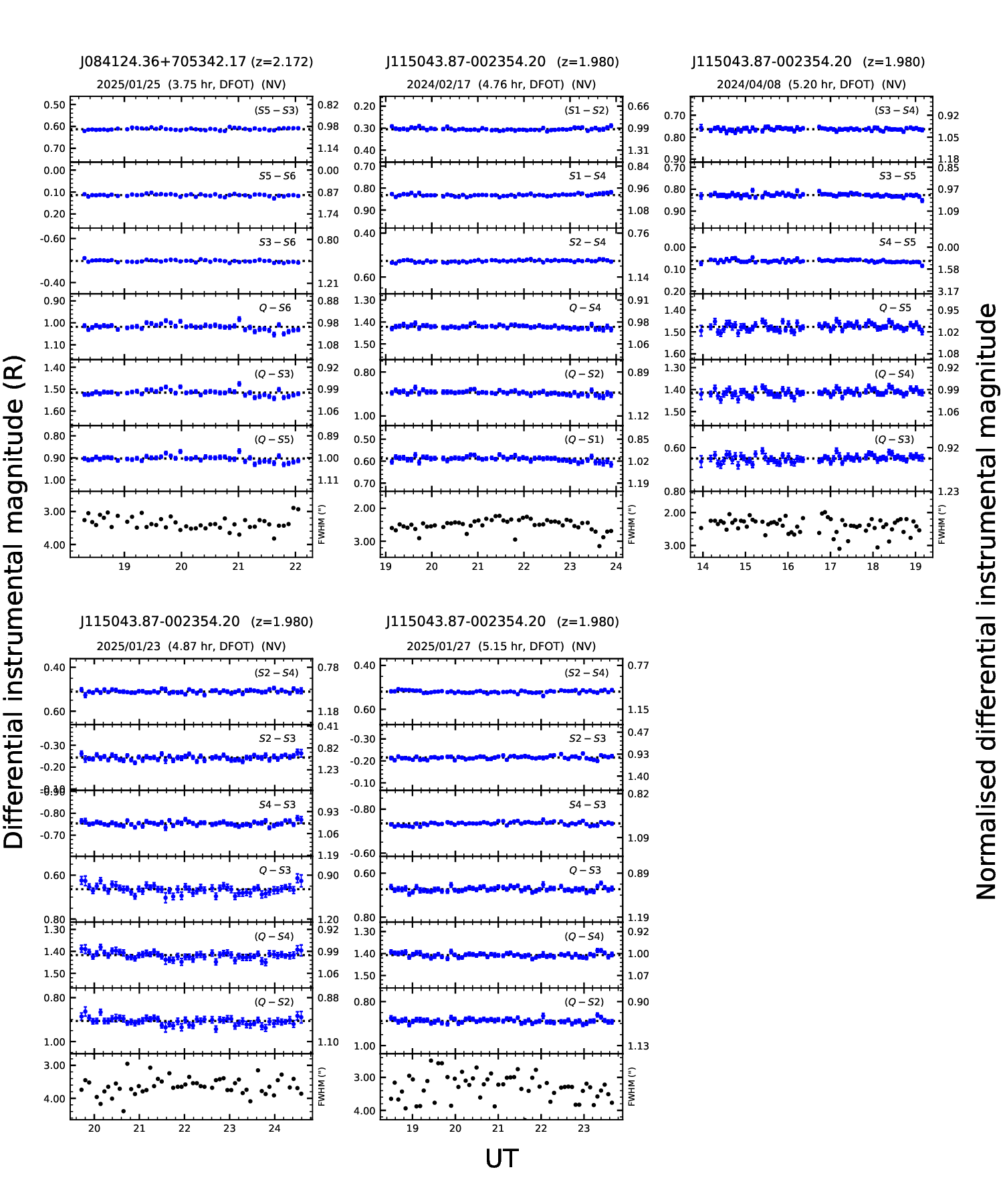}\\

  \vspace{-0.01in}
  \caption{Same as Fig.~\ref{fig1}.}
\label{fig2} 
\end{figure*}

\begin{figure*}

\includegraphics[width=1.08\textwidth,height=0.95\textheight]{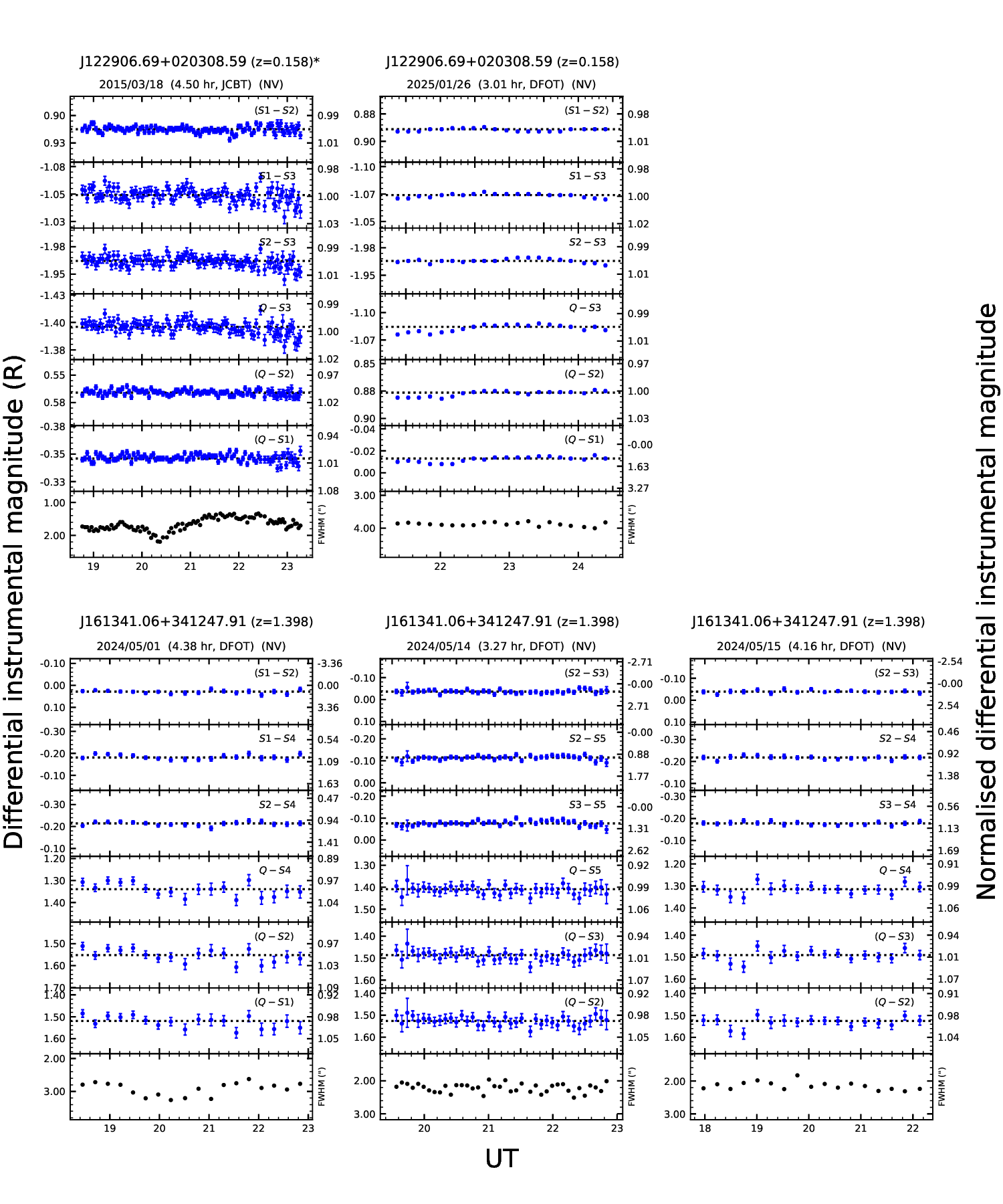}\\ 
  \vspace{-0.01in}
  \caption{Same as Fig.~\ref{fig1}.}
\label{fig3} 
\end{figure*}

\begin{figure*}

\includegraphics[width=1.08\textwidth,height=0.95\textheight]{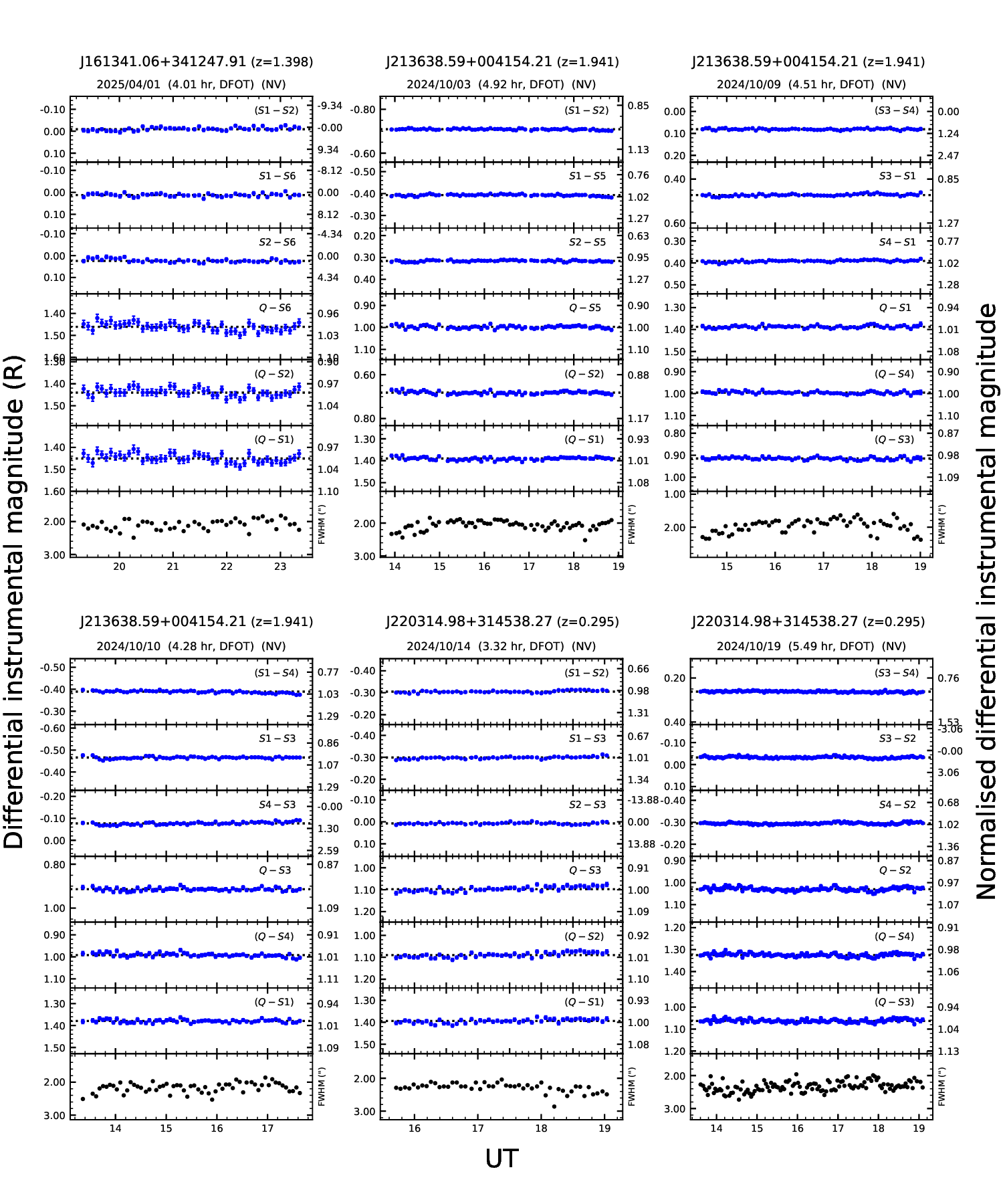}\\ 
  \vspace{-0.01in}
  \caption{Same as Fig.~\ref{fig1}.}
\label{fig4} 
\end{figure*}

\begin{figure*}

\includegraphics[width=0.45\textwidth,height=0.5\textheight]{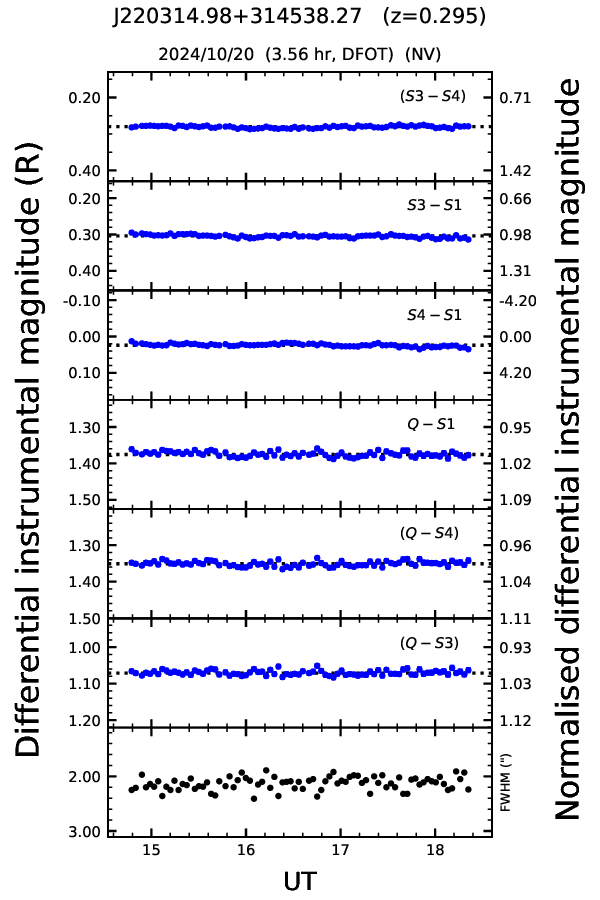}\\ 
  \vspace{-0.01in}
  \caption{Same as Fig.~\ref{fig1}.}
\label{fig5} 
\end{figure*}

\begin{figure*}
    \centering
        \begin{minipage}[t]{0.49\textwidth}
        \includegraphics[width=\textwidth]{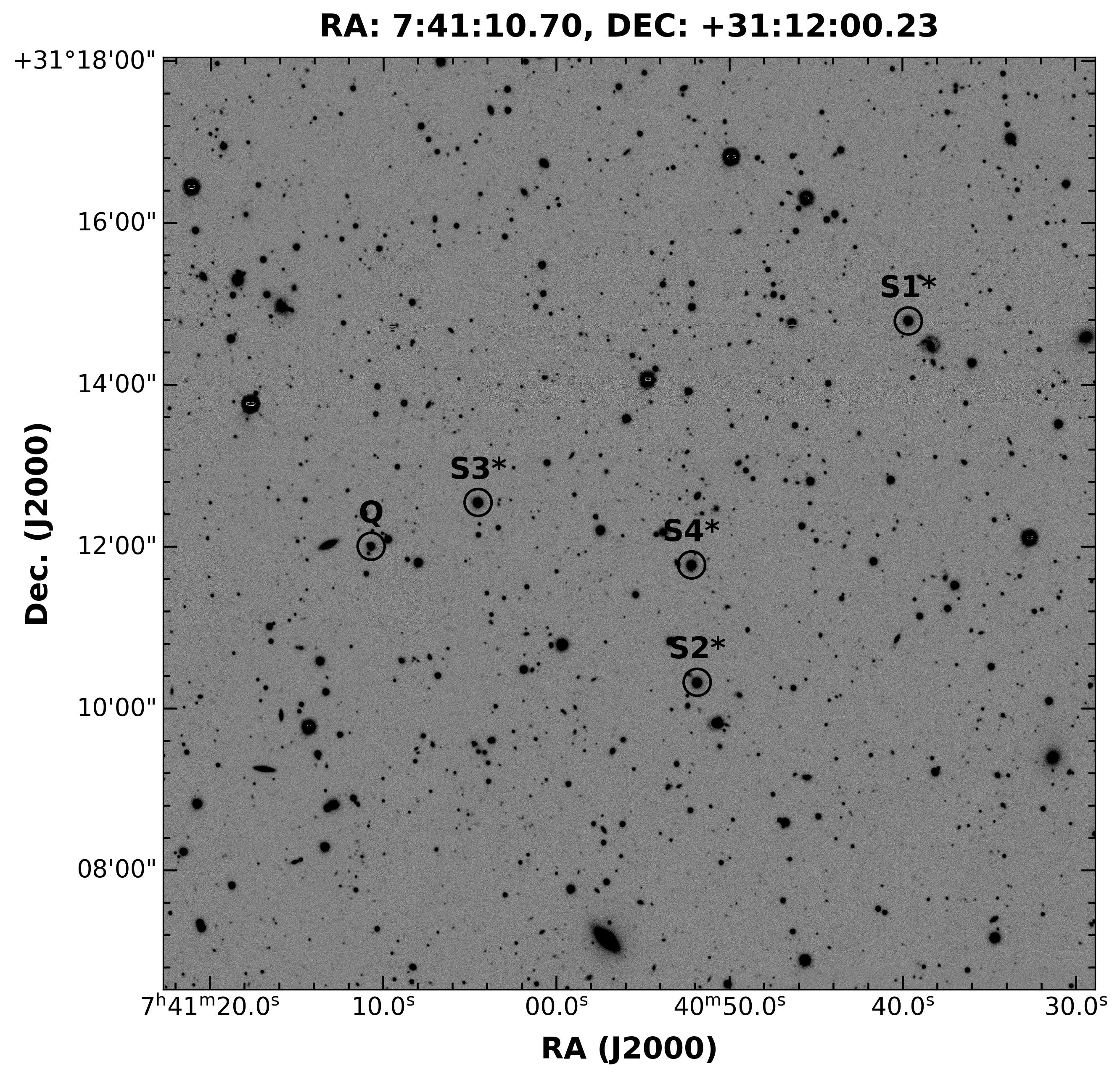}
        \caption{DECaLS finding chart for the target source J074110.70$+$311200.23 (Q) and the selected comparison stars (S1 to S4), indicated by circles. {Three comparison stars were utilized in each monitoring session for the $F_{enh}$ test, with the two steadiest stars selected for the $F_{\eta}$ test denoted by asterisks (*). Because the target field was observed across multiple sessions with slightly varying choice of comparison stars, more than three comparison stars are highlighted in the field.}}
        \label{fig:J0741_finding}
        \end{minipage}
    \hfill
    \begin{minipage}[t]{0.49\textwidth}
        \centering
        \includegraphics[width=\textwidth]{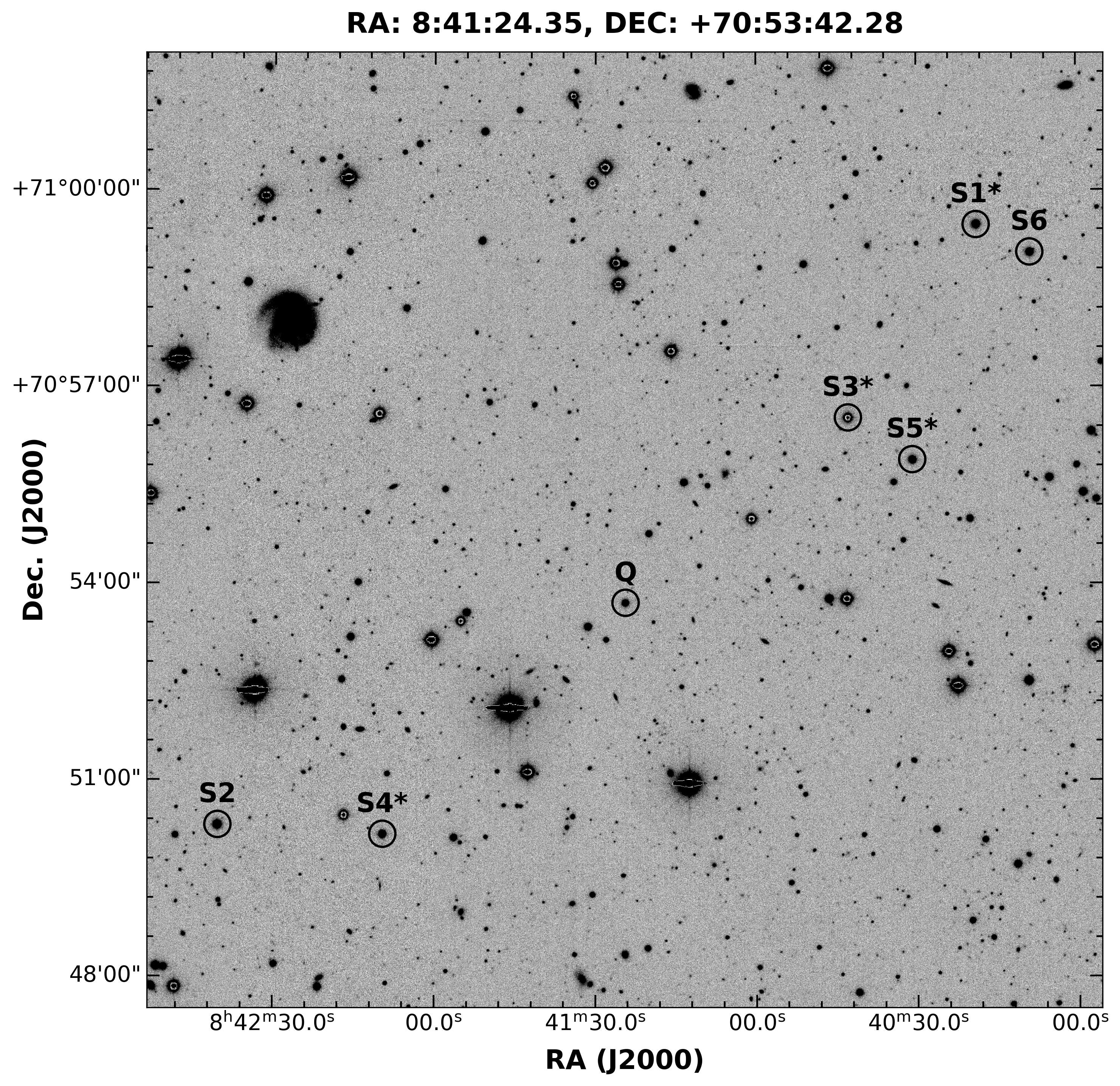}
        \caption{Same as Fig. \ref{fig:J0741_finding}, but for J084124.36$+$705342.17.}
        \label{fig:J0841_finding}
    \end{minipage}

\end{figure*}

Pre-processing of the raw CCD frames, including bias subtraction, flat-fielding and cosmic-ray removal, was performed using the standard tasks available in the Image Reduction and Analysis Facility (IRAF)\footnote{\url{http://iraf.noao.edu/}}. The instrumental magnitudes for the target sources and the carefully selected non-varying comparison stars registered in the same CCD frame were derived using aperture photometry \citep[] {Stetson1987PASP...99..191S,1992ASPC...25..297S}, implemented via the Dominion Astronomical Observatory Photometry II package (DAOPHOT II algorithm). For each frame, the `point spread function' (PSF) was estimated by taking the average of the full width at half maximum (FWHM) of the {Gaussians fitted to the} brightness profiles of five moderately bright stars within the CCD frame. An aperture radius of 2$\times$FWHM was adopted for the DAOPHOT photometry 
(see, e.g. \citealp{Krishan2022MNRAS.511L..13C}, also, \citealp{Vibhore2023MNRAS.524L..66N}). 
Differential light curves (DLCs) were then generated for each session, for all the pairs involving the target AGN and the selected three non-varying comparison stars. These DLCs, together with the plot of PSF variation during the session, are displayed in Fig. \ref{fig1} $-$ Fig. \ref{fig5}. {The observational parameters for the target sources and their corresponding comparison stars are presented in Table \ref{tab:comparison_color}. Column 1 gives the SDSS designation of the targets alongside their selected comparison stars. Column 2 lists the dates of the monitoring sessions. RA and Dec. are provided in columns 3 and 4, respectively. Columns 5 and 6 list the SDSS $g-$ and $r-$band magnitudes, while column 7 reports the derived $(g-r)$ colour index. Column 8 gives the colour index difference ($\Delta(g-r)$) between the target source and each associated comparison star. The colour dependent second-order atmospheric extinction across varying airmass during the course of monitoring can introduce colour dependent systematic trends into DLCs, potential mimicking spurious INOV. However, as investigated by \citet{Stalin2004JApA...25....1S} (see also, \citealp{Carini1992AJ....104...15C}), colour difference ($\Delta(g-r)$) between the target source and the comparison stars up to  2.0, or so, introduces negligible errors across different airmass ranges. Because the colour index difference between our target source and at least one comparison star is consistently $\Delta(g - r) < 1.5$ across our sample, second-order colour dependent atmospheric extinction effects are minimal and nowhere close to introducing spurious INOV. While SDSS $(g-r)$ colours were primary used for this assessment, SDSS photometry was unavailable for two targets in our sample: J084124.36$+$705342.17 and J220314.98$+$314538.27. For these two sources, we instead adopted $(B - R)$ colours from the USNO-A2.0 catalog (\citealp{Monet1998AAS...19312003M}). Optical finding charts for each of these 7 sources, along with the selected comparison stars, are provided in Fig. \ref{fig:J0741_finding} - Fig. \ref{fig:J22_finding}.} 

In order to raise the statistical robustness, we have augmented these analysis results with 19 intranight photometry sessions of duration > 3 hours {obtained using 1-metre class telescopes}, that were available in the literature for a part of our sample of 8 LPSQs. Note that in all these cases, the variation of PSF through the session was available for inspection, having been published along with the DLCs; in one instance involving {J122906.69$+$020308.59} (3C 273) (see below) the same could be determined from the digital data accessible to us. 
These additional 19 sessions consist of 4 sessions for J074110.70$+$311200.23 \citep{Goyal2012AA...544A..37G}; 2 sessions for J084124.36$+$705342.17 \citep{Krishan2022MNRAS.511L..13C}; {3 sessions for J122906.69$+$020308.59 (2 sessions from \citet{Pandian2022JApA...43...48P} and one session from \citet{Paliya2017ApJ...844...32P})}; 8 sessions for J135704.44$+$191907.37 \citep{Goyal2012AA...544A..37G} and 2 sessions for J220314.98$+$314538.27 \citep{Goyal2012AA...544A..37G} (see also, Table \ref{tab:comparison_color}). For 3 out of these 19 sessions, the monitoring data could be procured in digital form, namely,  J084124.36$+$705342.17 observed on 2019/12/17, 2020/02/14, and for J122906.69$+$020308.59 observed on 2015/03/18. This enabled us to re-analyse these data, thereby extending the uniformity of analysis procedure to these 3 sessions as well (see footnotes to the Table \ref{tab:comparison_color} for details, also, Figs. \ref{fig1} and \ref{fig3}). 
Out of the remaining 16 sessions, the INOV parameters were adopted from \citet{Goyal2012AA...544A..37G} for 14 sessions, and for the remaining 2 sessions targeting J122906.69$+$020308.59 (3C 273 at $z$ = 0.158), the INOV parameters could be taken from \citet{Pandian2022JApA...43...48P} because during these sessions PSF was reported to remain steady, thus fulfilling an important requirement for such relatively nearby quasars which are otherwise prone to spurious claims of INOV detection (see, e.g. \citealp{Cellone2000AJ....119.1534C}).

\begin{table*}
 \caption{Basic parameters of the comparison stars chosen for the monitored {target} quasars (LPSQs).
  \label{tab:comparison_color}}
\scriptsize
\begin{tabular}{ccc ccc r r}\\
\hline

{Target quasar and the} &   Date(s) of observations       &   {R.A. (J2000)} & {Dec. (J2000)}             & {\it g} & {\it r} & \hspace{0.10 cm} {\it g-r}  & { $\Delta(g-r)$} \\
comparison stars used
                  &  yyyy/mm/dd   &   (hh:mm:ss)       &($^\circ$: $^\prime$: $^{\prime\prime}$)   & (mag)   & (mag)   & (mag)  & Q-S (mag)     \\
{(1)}      & {(2)}        & {(3)}           & {(4)}                              & {(5)}   & {(6)}   &  \hspace{0.10 cm}{(7)}  &(8)   \\
\hline\\

J074110.70$+$311200.23 & 2006/01/20$^{G}$, 2006/01/21$^{G}$, 2006/12/18$^{G}$, 2006/12/22$^{G}$       & 07:41:10.70&$+$31:12:00.23 & 16.68 &  16.75 & $-$0.07 & ---  \\
              &2024/01/29, 2024/01/30, 2024/02/17, 2024/02/29 & & & &   &  &  \\
S1           &  2024/01/30, 2024/02/17, 2024/02/29 & 07:40:39.65&$+$31:14:47.29 & 16.25 &  15.42 & 0.83 & $-$0.90 \\
S2           & 2024/02/17, 2024/01/29, 2024/01/30, 2024/02/29             & 07:40:51.86&$+$31:10:19.44 & 15.49 &  15.17 & 0.32   & $-$0.39 \\
S3           & 2024/01/29, 2024/01/30, 2024/02/29             & 07:41:04.52&$+$31:12:32.79 & 15.79 &  15.27 & 0.52   & $-$0.59 \\
S4           & 2024/01/29, 2024/02/17             & 07:40:52.18&$+$31:11:46.46 & 15.76 &  15.24 & 0.52   & $-$0.59 \\
S1$^\prime$           & 2006/01/20, 2006/01/21, 2006/12/18, 2006/12/22 & 07:41:24.17&$+$31:09:44.52 & 15.61 &  15.26 & 0.35 & $-$0.42 \\
S2$^\prime$           & 2006/01/20, 2006/01/21, 2006/12/18, 2006/12/22 & 07:41:20.73&$+$31:08:49.56 & 15.81 &  15.38 & 0.43   & $-$0.50 \\
S3$^\prime$           & 2006/01/21, 2006/12/18, 2006/12/22             & 07:41:00.74&$+$31:16:44.77 & 16.86 &  16.33 & 0.53   & $-$0.60 \\
S4$^\prime$           & 2006/01/20            & 07:41:07.97&$+$31:11:48.39 & 16.30 &  15.87 & 0.43   & $-$0.50 \\\\

J084124.36$+$705342.17 & 2019/12/17$^C$, 2020/02/14$^{C}$, 2025/01/25    & 08:41:24.35&$+$70:53:42.28 &16.5  &16.7   &$-$0.20 &---  \\
  S1$^\dagger$         & 2019/12/17                           & 08:40:18.76&$+$70:59:28.37 &17.3  &15.1   &2.20   &$-$2.40 \\
  S2$^\dagger$         &  2019/12/17, 2020/02/14               & 08:42:40.23&$+$70:50:18.85 &15.8  &15.0   &0.80  & $-$1.00 \\
  S3$^\dagger$         &  2019/12/17, 2020/02/14, 2025/01/25 & 08:40:42.79&$+$70:56:31.72 &15.7  &15.1   &0.60   &$-$0.80\\
  S4$^\dagger$         &  2020/02/14                & 08:42:09.57&$+$70:50:10.46 &16.4  &15.5   &0.90   &$-$1.10\\
  S5$^\dagger$         &  2025/01/25                          & 08:40:30.80&$+$70:55:53.34 &17.5  &15.8   &1.70   &$-$1.90\\ 
  S6         &  2025/01/25                          & 08:40:08.74&$+$70:59:03.04 &16.1  &15.8   &0.30   &$-$0.50\\ \\

J115043.87$-$002354.20  & 2024/02/17, 2024/04/08, 2025/01/23, 2025/01/27 & 11:50:43.87&$-$00:23:54.20 &17.05  &17.00   & 0.05  & --- \\
                                                       
S1            & 2024/02/17 & 11:50:43.09&$-$00:29:02.60 &18.03  &16.55   & 1.48 & $-$1.43 \\
                                                      
S2            & 2024/02/17, 2025/01/23, 2025/01/27  & 11:50:13.02&$-$00:20:15.50 &17.01  &16.20   & 0.81  & $-$0.76 \\
S3            & 2024/04/08, 2025/01/23, 2025/01/27 & 11:50:52.57&$-$00:26:21.71 &16.78  &16.38   & 0.40 & $-$0.35 \\
                                                      
S4            & 2024/02/17, 2024/04/08, 2025/01/23, 2025/01/27  & 11:50:36.82&$-$00:21:22.06 &16.02  &15.64   & 0.38  & $-$0.33 \\
S5            & 2024/04/08 & 11:50:33.04&$-$00:15:19.26 &16.11  &15.61   & 0.50 & $-$0.45 \\\\

J122906.69$+$020308.59 & 2015/03/18$^P$, 2017/03/24$^{Pa}$, 2017/04/27$^{Pa}$, 2025/01/26   & 12:29:06.69&$+$02:03:08.59 & 13.00 &  12.89 & 0.11 & --- \\
S1$^\dagger$            & 2015/03/18, 2025/01/26 & 12:29:03.20&$+$02:03:18.60 &14.00  &13.36   & 0.64 & $-$0.53 \\
S2            & 2015/03/18, 2025/01/26 & 12:29:08.39&$+$02:00:18.88 &13.34  &14.04   & $-$0.70 & 0.81 \\
S3$^\dagger$            & 2015/03/18, 2025/01/26 & 12:29:02.83&$+$02:02:16.14 &15.49  &14.46   & 1.03 & $-$0.92 \\
S1$^\prime$            & 2017/03/24, 2017/04/27                & 12:29:03.20&$+$02:03:18.60 &14.00  &13.36   & 0.64 & $-$0.53 \\
S2$^\prime$            & 2017/03/24, 2017/04/27                & 12:29:02.83&$+$02:02:16.14 &15.49  &14.46   & 1.03 & $-$0.92 \\
S3$^\prime$            & 2017/03/24, 2017/04/27                & 12:29:07.87&$+$02:03:35.93 &17.76  &16.27   & 1.49 & $-$1.38 \\\\

J135704.44$+$191907.37 & 2006/02/27$^{G}$, 2006/03/05$^{G}$, 2006/03/26$^{G}$, 2006/03/28$^{G}$,    & 13:57:04.44&$+$19:19:07.37 & 16.17 &  16.09 & 0.08 & --- \\
           &2006/03/29$^{G}$, 2006/04/06$^{G}$, 2006/04/22$^{G}$, 2006/04/23$^{G}$  & & & &   &  &  \\
S1$^\prime$            &2006/02/27, 2006/03/05, 2006/03/26, 2006/03/28,   & 13:57:04.60&$+$19:20:24.20 &16.60  &15.96   & 0.64 & $-$0.56 \\
              &2006/03/29, 2006/04/06, 2006/04/22, 2006/04/23  & & & &   &  &  \\
S2$^\prime$            &2006/02/27, 2006/03/05, 2006/03/26, 2006/03/28,   & 13:57:07.00&$+$19:22:30.00 &17.67  &16.22   & 1.45 & $-$1.37 \\
              &2006/03/29, 2006/04/06, 2006/04/22  & & & &   &  &  \\
S3$^\prime$            &2006/02/27, 2006/03/26, 2006/03/28, 2006/03/29,   & 13:57:19.36&$+$19:17:57.70 &17.24  &16.08   & 1.16 & $-$1.08 \\
              &2006/04/22, 2006/04/23  & & & &   &  &  \\
S4$^\prime$            &2006/03/05, 2006/04/06  & 13:56:52.13&$+$19:20:51.80 &16.01  &15.24   & 0.77 & $-$0.69 \\
S5$^\prime$            &2006/04/23  & 13:56:52.79&$+$19:14:59.20 &16.11  &15.47   & 0.64 & $-$0.56 \\\\

J161341.06$+$341247.91 & 2024/05/01, 2024/05/14, 2024/05/15, 2025/04/01   & 16:13:41.06&$+$34:12:47.91 & 17.63 &  17.32 & 0.31  & --- \\
                                 
S1           & 2024/05/01, 2025/04/01    & 16:14:15.18&$+$34:08:27.01 & 17.26 &  16.74 & 0.52  &  $-$0.21\\
                                 
S2           & 2024/05/01, 2024/05/14, 2024/05/15, 2025/04/01     & 16:14:05.69&$+$34:19:12.79 & 17.58 &  16.75 & 0.83   & $-$0.52 \\
S3           & 2024/05/01, 2024/05/14, 2024/05/15    & 16:14:27.46&$+$34:06:08.67 & 17.62 &  16.97 & 0.65   & $-$0.34\\
S4           & 2024/05/14, 2024/05/15    & 16:14:10.63&$+$34:13:40.84 & 17.85 &  16.80 & 1.05   & $-$0.74\\
S5           & 2024/05/14    & 16:14:17.89&$+$34:14:08.63 & 17.48 &  16.86 & 0.62   & $-$0.31\\
S6           & 2025/04/01   & 16:14:05.37&$+$34:14:51.24 & 17.59 &  16.69 & 0.90   & $-$0.59\\\\

J213638.59$+$004154.21 &  2024/10/03, 2024/10/09, 2024/10/10 & 21:36:38.59 &$+$00:41:54.22 & 16.91 &  16.84 & 0.07  & --- \\
 S1 	      &  2024/10/03, 2024/10/09, 2024/10/10   & 21:36:26.22 &$+$00:43:41.28 & 15.76 &  15.26 & 0.50   & $-$0.43 \\
 S2 	      &  2024/10/03              & 21:36:45.90 &$+$00:49:12.62 & 16.68 &  16.02 & 0.66   & $-$0.59\\
 S3           &  2024/10/09, 2024/10/10            & 21:36:31.19 &$+$00:48:39.90 & 16.35 &  15.78 & 0.57  & $-$0.50 \\
 S4           &  2024/10/09, 2024/10/10             & 21:36:29.40 &$+$00:47:48.09 & 17.13 &  15.75 & 1.38 & $-$1.31\\
 S5           &  2024/10/03              & 21:36:29.62 &$+$00:45:02.27 & 16.91 &  15.70 & 1.21   & $-$1.14\\\\

J220314.98$+$314538.27 & 2006/09/14$^{G}$, 2007/09/15$^{G}$, 2024/10/14, 2024/10/19, 2024/10/20 & 22:03:14.98&$+$31:45:38.27 & 15.70 &  14.40 & 1.30   &--- \\
                                                     
S1           & 2024/10/14, 2024/10/20             & 22:03:10.93&$+$31:43:54.73 & 15.30 &  13.90 & 1.40   & $-$0.10\\
                                                      
S2           & 2024/10/14, 2024/10/19             & 22:03:28.34&$+$31:42:10.20 & 15.70 &  14.60 & 1.10   & 0.20\\
                                                      
S3           & 2024/10/14, 2024/10/19, 2024/10/20 & 22:03:24.19&$+$31:43:26.13 & 15.30 &  14.50 & 0.80  & 0.50 \\

S4           & 2024/10/19, 2024/10/20 & 22:03:35.98&$+$31:41:05.95 & 16.00 &  14.30 & 1.70   & $-$0.40 \\

S1$^\prime$           & 2006/09/14, 2007/09/15             & 22:02:58.00&$+$31:48:43.30 & 16.00 &  15.05 & 0.95   & 0.35\\
                                                      
S2$^\prime$           & 2006/09/14, 2007/09/15             & 22:03:27.10&$+$31:41:47.40 &  15.86 & 14.64   & 1.22 & 0.08\\
                                                      
S3$^\prime$           & 2006/09/14, 2007/09/15 & 22:02:52.30&$+$31:46:51.20 & 15.75 &  15.03 & 0.72  & 0.58 \\\\

\hline\\

\multicolumn{8}{l}{Intranight monitoring data for four sessions targeting the quasar J074110.70$+$311200.23 (marked with the letter, `$^G$') were adopted from \citet{Goyal2012AA...544A..37G}. Two sessions for }\\
\multicolumn{8}{l}{J084124.36+705342.17 (marked with the letter, `$^C$'), one session for J122906.69$+$020308.59 (marked with the letter, `$^P$'), two sessions for J122906.69$+$020308.59 (marked with  }\\
\multicolumn{8}{l}{the letter, `$^{Pa}$'), all eight sessions for J135704.44$+$191907.37, and two sessions for J220314.98$+$314538.27 (marked with the letter, `$^G$') were adopted from \citet{Krishan2022MNRAS.511L..13C}, }\\
\multicolumn{8}{l}{ \citet{Paliya2017ApJ...844...32P}, \citet{Pandian2022JApA...43...48P}, and \citet{Goyal2012AA...544A..37G}, respectively. {Comparison stars used in these literature sessions are indicated with a prime symbol ($^\prime$). }}\\
\multicolumn{8}{l}{{$\dagger$ Comparison stars utilized in the present study were adopted from previous INOV studies: \citet{Krishan2022MNRAS.511L..13C} for J084124.36$+$705342.17 and \citet{Paliya2017ApJ...844...32P} for }}\\
\multicolumn{8}{l}{J122906.69$+$020308.59.}

\end{tabular}

\end{table*}

\section{Statistical Analysis and Results}
\label{stat}
{The statistical analysis presented in this section was performed on the DLCs obtained in our 20 monitoring sessions and another 3 sessions for which digital data also became available to us (see above). The Johnson-Cousins $R$-band was used in all the 23 sessions. To search for INOV in all
these DLCs, we employed the widely used $F_{\eta}$ statistical test \citep{Diego2010AJ....139.1269D}, supplemented with another test, the enhanced F-test, ($F_{enh}$; \citealp{Diego2014AJ....148...93D}).}
\subsection{Application of the $F_{\eta}$ Test}
\label{f-eta}
Details of the procedure followed for this test are provided in \citet{Gopal-krishna2023MNRAS.518L..13G}, \citet{Vibhore2023MNRAS.524L..66N} and \citet{Krishan2024PASA...41..106C}. {For each observing session, three candidate comparison stars were initially chosen. Out of these, the two most stable stars were identified by inspecting all the `star–star' DLCs. The $F_{\eta}$ test was applied exclusively to the DLCs derived using these two selected comparison stars and the target source (see also, \citealp{Krishan2022MNRAS.511L..13C}). 
The labels of these selected DLCs are shown within parentheses in the corresponding panels (see Fig. \ref{fig1} - Fig. \ref{fig5}).} Several independent studies have reported that the photometric errors provided by DAOPHOT are systematically underestimated by a factor $\eta$, typically ranging between 1.30 and 1.75. Ignoring this correction may often lead to spurious claims of INOV detection \citep{Gopal-Krishna1995MNRAS.274..701G,Garcia1999MNRAS.309..803G,Sagar2004MNRAS.348..176S,Stalin2004MNRAS.350..175S,Bachev2005MNRAS.358..774B,Goyal2012AA...544A..37G,Goyal2013MNRAS.435.1300G}. Therefore, we have adopted here $\eta$ = 1.54 estimated to $3\%$ accuracy in a detailed analysis of 262 DLCs involving pairs of non-varying comparison stars \citep{Goyal2013JApA...34..273G}. 
For each monitoring session, the $F^{\eta}$ values were computed for the two DLCs of the target AGN (relative to the two chosen comparison stars), as:
\begin{equation} 
\label{eq.ftest2}
F_{1}^{\eta} = \frac{Var(q-s1)}
{ \eta^2 \sum_\mathbf{i=1}^{N}\sigma^2_{i,err}(q-s1)/N},  \\
\hspace{0.1cm} F_{2}^{\eta} = \frac{Var(q-s2)}
{ \eta^2 \sum_\mathbf{i=1}^{N}\sigma^2_{i,err}(q-s2)/N}  \\
\end{equation}\\
Here, $Var(q-s1)$ and $Var(q-s2)$ represent the variances of the target's DLCs relative to the two comparison stars, while $\sigma_{i,err}(q-s1)$ and $\sigma_{i,err}(q-s2)$ denote the rms errors returned by DAOPHOT for the $i^{th}$ data point in the corresponding target's DLCs.
$N$ is the number of data points in the DLCs of the session, and $\eta = 1.54$ is the scaling factor used to correct for the underestimation of the photometric errors, as mentioned above. {For each session, in addition to computing $F_{1}^{\eta}$ and $F_{2}^{\eta}$ for the target's DLCs relative to the two comparison stars, we also computed the critical values of $F $ ($= F_{c}^{\alpha}$), for $\alpha = $ 0.05 and 0.01, corresponding to confidence levels for INOV detection of 95\% and 99\%, respectively.} These critical values are used to assess the statistical significance of INOV  by comparing them with the $F^{\eta}$ values computed for the two DLCs of the target, using Eq. (\ref{eq.ftest2}), namely, $F_{1,2}^{\eta}$.  
The target source is classified as `variable' (V) in a session if both its computed $F^{\eta}$ values exceed $\ge$ $F_{c}$(0.99).  
The `Photometric Noise Parameter' for each session is defined as PNP = {$\sqrt { \eta^2\langle \sigma^2_{i,err} \rangle}$ }, where $\eta=1.54$.

\subsection{Application of the Enhanced F-test}
{We have augmented the statistical checking for INOV by subjecting the DLCs for the above 23 sessions to `enhanced F-test' ($F_{enh}$), which evaluates the flux variability by comparing the variance of the target's DLC with the stacked variances of the comparison stars{, all derived relative to a same reference star} (e.g. \citealp{Polednikova2016MNRAS.460.3950P}; \citealp{Bhattacharya2019MNRAS.483.3382B}; \citealp{Sapna2021MNRAS.507L..46M}). A particular advantage of the $F_{enh}$ test is that it corrects for any magnitude differences between the target and the comparison star(s), by appropriately scaling the DLCs of the comparison stars. This scaling adjusts the photometric noise of the comparison stars {by making} their {respective} magnitudes {match} the mean magnitude of the target source. Furthermore, incorporating multiple comparison stars enhances the statistical power of the test. Among the available comparison stars, the one closest in magnitude to the target source is selected as the reference star (\citealp{Bhattacharya2019MNRAS.483.3382B}; \citealp{Devanand2025ApJ...990..214D}). {For each session, as mentioned above, three comparison stars were selected and used for the $F_{enh}$ test, one of them as the reference star. From this same set, the two stars found to be most stable were utilized for the $F_{\eta}$ test (Section \ref{f-eta}),  
rendering them common to both statistical tests performed here, namely $F_{\eta}$ and $F_{enh}$ tests.} 
The statistical criterion for the `enhanced F-test' is defined as:

\begin{equation}
\label{F-enh}
    \begin{aligned}
F_{{enh}} &=
\frac{{Var}(q-{ref})}{{Var}_{c}},
&
{Var}_{c} &=
\frac{1}{\displaystyle\sum_{j=1}^{k}N_j-k}
\sum_{j=1}^{k}
\sum_{i=1}^{N_j}
s_{j,i}^{2}
\end{aligned}
\end{equation}

where $Var(q-ref)$ represents the variance of the target with respect to reference star, and $Var_c$ denotes the stacked variance of the comparison stars with respect to reference star. The term $N_j$ denotes the number of observations for the $j^{th}$ comparison star, $k$ is the total number of comparison stars, and $s_{j,i}^{2}$ represents the scaled square deviation, defined as:

\begin{equation}
s_{j,i}^{2}
=
\omega_j
\left(m_{j,i}-\overline{m}{_j}\right)^2
\end{equation}

Here, $m_{j,i}$'s and $\overline{m}{_j}$ correspond to the $i^{th}$ differential magnitude and the mean differential magnitude of the $j^{th}$ comparison-reference star DLC, respectively. The scaling factor $\omega_j$ is defined as:

\begin{equation}
\omega_j =
\frac{
\left\langle
\sigma_{{err}}^{2}(q-{ref})
\right\rangle
}{
\left\langle
\sigma_{{err}}^{2}(s_j-{ref})
\right\rangle
}
\end{equation}

where $\left\langle\sigma_{{err}}^{2}(q-{ref})\right\rangle$ and $\left\langle\sigma_{{err}}^{2}(s_j-{ref})\right\rangle$ represent the session-averaged variances of the target-reference star and the $j^{th}$ comparison star-reference star DLCs, respectively, derived from the individual photometric rms errors ($\sigma_{{err}}$) returned by the DAOPHOT. Because the comparison stars and the target are observed simultaneously within the same CCD frame, the number of observations is identical for all sources ($N_j = N$). Consequently, the $F_{{enh}}$ statistic (Eq. \ref{F-enh}) follows an $F$-distribution with numerator and denominator degrees of freedom (DOF) of $\nu_1=N-1$ and $\nu_2=k(N-1)$, respectively.

For each target DLC, we computed the value of $F_{enh}$ and compared it against the critical value, F $(F^\alpha_c(\nu_1,\nu_2))$ evaluated for $\alpha$
= 0.01, corresponding to a 99\% confidence level for INOV detection. If the computed value of $F_{enh} \geq F_c (0.99)$ for a DLC, the null hypothesis (i.e. no variability) is rejected, and the target is designated as variable (`V') at the $\geq$ 99\% confidence level.}

\subsection{Variability Amplitude}
The INOV variability amplitude ($\psi$) was estimated following \citet{Heidt1996A&A...305...42H}:\\
 $\psi= \sqrt{({A_{max}}-{A_{min}})^2-2\sigma^2}$\\
where $A_{max}$ and $A_{min}$ are the maximum and minimum values in the target-comparison star DLC and $\sigma^2=\eta^2<\sigma^2_{q-s}>$, with $\sigma^2_{q-s}$ being the mean square rms error for the data points in the DLC and ${\eta=1.54}$. For each session classified as variable, the mean value of $\psi$ is computed as the average of the $\psi$ values estimated from the two DLCs of the target AGN, relative to the two comparison stars.

{We note that even though, the above formulation of variability amplitude ($\psi$) was originally conceptualized for flux variability of a source (light-curve), its adaptation to DLCs has become a standard practice in INOV studies (e.g. \citealp{Romero1999A&AS..135..477R}, \citealp{GopalKrishna2003ApJ...586L..25G}, \citealp{Goyal2012AA...544A..37G}, \citealp{Polednikova2015A&A...578A.121P}). Because differential photometry effectively cancels out any atmospheric and instrumental fluctuations during the session, small variation ($\Delta m$) in the DLC can be directly mapped to fractional flux changes through a given session, justifying the use of the above definition of $\psi$ even in the case of DLCs.}

\subsection{Duty Cycle}
The duty cycle (DC) of INOV was evaluated using the definition of \citet{Romero1999A&AS..135..477R}:
\begin{equation} 
DC = 100\frac{\sum_\mathbf{i=1}^\mathbf{n} N_i(1/\Delta T_i)}{\sum_\mathbf{i=1}^\mathbf{n}(1/\Delta T_i)} {\rm \%} 
\label{eq:dc} 
\end{equation}
 where $\Delta T_i = \Delta T_{i,obs}(1+z)^{-1}$ is the intrinsic duration of the $i^{th}$ session, corrected for redshift, $z$ of the source and $\Delta T_{i,obs}$ is the observed duration. $N_i$ = 1 if variability is detected in the $i^{th}$ session, otherwise $N_i$ = 0. 

 \renewcommand{\thetable}{3a}
\begin{table*}
\centering
\scriptsize
\setlength{\tabcolsep}{2.5pt} 
\renewcommand{\arraystretch}{1.15}

\caption{Results of the statistical tests $F_{\eta}$ and $F_{\rm enh}$ for detecting INOV in the DLCs of the low-polarization superluminal quasars (LPSQs) ($\eta$ = 1.54).}
\label{tab:combined_inov_stats}

\resizebox{\textwidth}{!}{%
\begin{tabular}{@{} l c c c c c c c c c c c c c @{}}
\toprule \toprule
\multicolumn{1}{c}{Quasar} & Obs. date & N & T & \multicolumn{3}{c}{$F_{\eta}$ Test} & \multicolumn{2}{c}{INOV Status$^{\rm a}$} & \multicolumn{3}{c}{Enhanced F-test ($F_{\rm enh}$)} & $\sqrt{\eta^2 \langle \sigma^2_{i,{\rm err}} \rangle}$ & $\overline{\psi}$$^{\rm b}$ \\
\cmidrule(lr){5-7} \cmidrule(lr){8-9} \cmidrule(lr){10-12}
\multicolumn{1}{c}{(SDSS Name)} & yyyy/mm/dd & & (hr) & $F_1^\eta, F_2^\eta$ & $F_{\rm c}(0.95)$ & $F_{\rm c}(0.99)$ & Using star1, star2 & {Adopted} & $F_{\rm enh}$ & $F_{\rm c}(0.99)$ & Status & & (\%) \\
\multicolumn{1}{c}{(1)} & (2) & (3) & (4) & (5) & (6) & (7) & (8) & (9) & (10) & (11) & (12) & (13) & (14) \\
\midrule \midrule
J074110.70+311200.23 & 2024/01/29 & 50 & 4.47 & 0.57 (S4), 0.53 (S3) & 1.61 & 1.96 & NV (S4), NV (S3) & NV & 0.42 & 1.74 & NV & 0.006 & --- \\
J074110.70+311200.23 & 2024/01/30 & 52 & 4.42 & 1.49 (S1), 1.04 (S3) & 1.59 & 1.94 & NV (S1), NV (S3) & NV & 1.82 & 1.73 & V  & 0.005 & 3.77 \\
J074110.70+311200.23 & 2024/02/17 & 55 & 4.69 & 0.89 (S1), 0.63 (S2) & 1.57 & 1.90 & NV (S1), NV (S2) & NV & 1.14 & 1.70 & NV & 0.006 & --- \\
J074110.70+311200.23 & 2024/02/29 & 61 & 4.35 & 1.04 (S1), 0.60 (S3) & 1.53 & 1.84 & NV (S1), NV (S3) & NV & 0.74 & 1.66 & NV & 0.004 & --- \\[1ex]

J084124.36+705342.17$^{C}$ & 2019/12/17 & 42 & 3.54 & 0.60 (S1), 0.77 (S3) & 1.68 & 2.09 & NV (S1), NV (S3) & NV & 0.59 & 1.84 & NV & 0.007 & --- \\
J084124.36+705342.17$^{C}$ & 2020/02/14 & 49 & 4.27 & 0.72 (S4), 0.60 (S3) & 1.62 & 1.98 & NV (S4), NV (S3) & NV & 0.77 & 1.75 & NV & 0.007 & --- \\
J084124.36+705342.17       & 2025/01/25 & 45 & 3.75 & 1.09 (S5), 1.27 (S3) & 1.65 & 2.04 & NV (S5), NV (S3) & NV & 1.93 & 1.80 & V  & 0.006 & 6.12 \\[1ex]

J115043.87$-$002354.20 & 2024/02/17 & 55 & 4.76 & 0.56 (S1), 0.44 (S2) & 1.57 & 1.90 & NV (S1), NV (S2) & NV & 1.29 & 1.70 & NV & 0.009 & --- \\
J115043.87$-$002354.20 & 2024/04/08 & 67 & 5.20 & 0.54 (S3), 0.58 (S4) & 1.50 & 1.78 & NV (S3), NV (S4) & NV & 1.52 & 1.62 & NV & 0.011 & --- \\
J115043.87$-$002354.20 & 2025/01/23 & 57 & 4.87 & 0.54 (S2), 0.55 (S4) & 1.56 & 1.88 & NV (S2), NV (S4) & NV & 1.32 & 1.68 & NV & 0.011 & --- \\
J115043.87$-$002354.20 & 2025/01/27 & 59 & 5.15 & 0.44 (S2), 0.50 (S4) & 1.55 & 1.86 & NV (S2), NV (S4) & NV & 0.61 & 1.67 & NV & 0.007 & --- \\[1ex]

J122906.69+020308.59$^{P}$ & 2015/03/18 & 97 & 4.50 & 0.75 (S1), 0.51 (S2) & 1.40 & 1.61 & NV (S1), NV (S2) & NV & 1.03 & 1.49 & NV & 0.003 & --- \\
J122906.69+020308.59       & 2025/01/26 & 20 & 3.01 & 2.52 (S1), 2.64 (S2) & 2.17 & 3.03 & NV (S1), NV (S2) & NV & 2.33 & 2.42 & NV & 0.002 & --- \\[1ex]

J161341.06+341247.91 & 2024/05/01 & 18 & 4.38 & 0.57 (S1), 0.71 (S2) & 2.27 & 3.24 & NV (S1), NV (S2) & NV & 1.45 & 2.54 & NV & 0.012 & --- \\
J161341.06+341247.91 & 2024/05/14 & 39 & 3.27 & 0.22 (S2), 0.24 (S3) & 1.72 & 2.16 & NV (S2), NV (S3) & NV & 0.47 & 1.88 & NV & 0.015 & --- \\
J161341.06+341247.91 & 2024/05/15 & 17 & 4.16 & 0.44 (S2), 0.51 (S3) & 2.33 & 3.37 & NV (S2), NV (S3) & NV & 1.34 & 2.62 & NV & 0.012 & --- \\
J161341.06+341247.91 & 2025/04/01 & 48 & 4.01 & 0.69 (S1), 0.58 (S2) & 1.62 & 1.99 & NV (S1), NV (S2) & NV & 0.85 & 1.76 & NV & 0.009 & --- \\[1ex]

J213638.59+004154.21 & 2024/10/03 & 68 & 4.92 & 0.61 (S1), 0.50 (S2) & 1.50 & 1.78 & NV (S1), NV (S2) & NV & 1.41 & 1.61 & NV & 0.006 & --- \\
J213638.59+004154.21 & 2024/10/09 & 66 & 4.51 & 0.48 (S3), 0.43 (S4) & 1.51 & 1.79 & NV (S3), NV (S4) & NV & 0.71 & 1.62 & NV & 0.006 & --- \\
J213638.59+004154.21 & 2024/10/10 & 61 & 4.28 & 0.52 (S1), 0.71 (S4) & 1.53 & 1.84 & NV (S1), NV (S4) & NV & 0.55 & 1.66 & NV & 0.005 & --- \\[1ex]

J220314.98+314538.27 & 2024/10/14 & 49  & 3.32 & 0.75 (S1), 1.04 (S2) & 1.62 & 1.98 & NV (S1), NV (S2) & NV & 1.41 & 1.75 & NV & 0.005 & --- \\
J220314.98+314538.27 & 2024/10/19 & 125 & 5.49 & 0.75 (S3), 0.83 (S4) & 1.35 & 1.52 & NV (S3), NV (S4) & NV & 1.05 & 1.42 & NV & 0.004 & --- \\
J220314.98+314538.27 & 2024/10/20 & 82  & 3.56 & 0.86 (S3), 0.91 (S4) & 1.44 & 1.68 & NV (S3), NV (S4) & NV & 1.18 & 1.55 & NV & 0.004 & --- \\
\bottomrule
\end{tabular}%
}

\smallskip
\begin{minipage}{\textwidth}
$^{a}$ V = variable, i.e., confidence $\geq 0.99$; NV = non-variable ($< 0.99$).\\
Variability status identifiers (col. 8), based on Quasar-star1 and Quasar-star2 DLCs are separated by a comma.\\
{$^{b}$ The reported variability amplitude (${\overline\psi}$) is computed for sessions with INOV detection using at least one statistical test ($F_{\eta}$ or $F_{\rm {enh}}$). Because all the sessions were classified as `Non-Variable' (NV) using the standard $F_{\eta}$ test, the quoted $\overline{\psi}$ values correspond only to sessions with INOV detection (type `V') using the enhanced F-test ($F_{\rm enh}$).}\\
The INOV data for two sessions of J084124.36+705342.17, marked with a letter (${C}$) were adopted from \citet{Krishan2022MNRAS.511L..13C} and re-analysed in the present study.\\
One INOV session of J122906.69+020308.59, marked with a letter (${P}$) was adopted from \citet{Paliya2017ApJ...844...32P} and those data have been re-analysed in the present study.
\end{minipage}
\end{table*}

\renewcommand{\thetable}{3b}
\begin{table}
 \centering
  
\caption{Additional INOV results (taken directly from the literature) for a part of our sample of low-polarization superluminal quasars (LPSQs).}
\label{tab2:3c}
\setlength{\tabcolsep}{3.5pt}
\begin{tabular}{ccccccc}
  \hline \hline
\multicolumn{1}{c}{Quasar}
&\multicolumn{1}{c}{Obs. date}
&\multicolumn{1}{c}{Tel.$^{a}$}
&{N}
&{T} 
 &\multicolumn{1}{c}{INOV}
 
  & $\overline\psi$ \\
  (SDSS name)       & yyyy/mm/dd & used & &{(hr)} &\multicolumn{1}{c}{status} &\%\\
 (1)     &   (2)    &(3)&(4)  & (5)                         &(6)       &(7)                \\
\hline\hline\\

J074110.70+311200.23$^{G}$ & 2006/01/20 & ST &29 &  7.00  & NV       &  ---       \\
J074110.70+311200.23$^{G}$ & 2006/01/21 & ST &16 &  3.60  & V        &  4.88  \\
J074110.70+311200.23$^{G}$ & 2006/12/18 & ST &28 &  6.80  & NV       &  --- \\
J074110.70+311200.23$^{G}$ & 2006/12/22 & ST &30 &  7.30  & V        &  1.33 \\\\

J122906.69+020308.59$^{Pa}$ & 2017/03/24 & JCBT &65   &  6.24&      V              &  1.90 \\
J122906.69+020308.59$^{Pa}$ & 2017/04/27 & JCBT &69   &  4.50&      V                &  4.04  \\\\

J135704.44$+$191907.37$^{G}$ & 2006/02/27 & ST &10  &  4.20&       V       &  1.25  \\
J135704.44$+$191907.37$^{G}$ & 2006/03/05 & ST &09   &  4.00&       NV       &  ---  \\
J135704.44$+$191907.37$^{G}$ & 2006/03/26 & ST &10   &  5.80&       NV       &  ---  \\
J135704.44$+$191907.37$^{G}$ &  2006/03/28 & ST &18   &  5.20&      V       &  3.58  \\
J135704.44$+$191907.37$^{G}$ & 2006/03/29 & ST &19   &  5.30&      NV       &  ---  \\
J135704.44$+$191907.37$^{G}$ & 2006/04/06 & ST &24   &  6.80&       NV       &  ---  \\
J135704.44$+$191907.37$^{G}$ & 2006/04/22 & ST &15   &  4.10&      NV       &  ---  \\
J135704.44$+$191907.37$^{G}$ & 2006/04/23 & ST &14   &  4.40&       V       &  2.16  \\\\

J220314.98+314538.27$^{G}$ & 2006/09/14 & ST &24  &  5.40&      V      &  0.90 \\
J220314.98+314538.27$^{G}$ & 2007/09/15 & ST &30  &  7.10&      NV     &  --- \\\\

\hline
\end{tabular}

\vspace{2mm}

\begin{minipage}{\textwidth}
\footnotesize

$^a$ ST – Sampurnanand Telescope; JCBT - J. C. Bhattacharya Telescope.\\
Ref. codes are: G (\citealp{Goyal2012AA...544A..37G}) and Pa (\citealp{Pandian2022JApA...43...48P}).

 \end{minipage}
\end{table}

\section{Results and discussion}

Fig. \ref{fig1} $-$ Fig. \ref{fig5} display the differential light curves (DLCs) for our sample of LPSQs (Table \ref{tab:Table1}), all using the 1.3-metre telescopes (22 sessions using DFOT and one session using JCBT\footnote{\url{https://www.iiap.res.in/centers/vbo/jcbt/}}). These include {20} DLCs being reported for the first time, {2} DLCs (for J084124.36$+$705342.17) 
from \citet{Krishan2022MNRAS.511L..13C} and one DLC for 3C 273 (J122906.69$+$020308.59) which had been monitored by \citet{Paliya2017ApJ...844...32P} on 2015/03/18 using the 1.3-metre JCBT (since the INOV data for the last 3 sessions were also available to us in digital form, we have analysed them anew, in order to maintain uniformity in the analysis). INOV results based on all these 23 DLCs are presented in Table \ref{tab:combined_inov_stats}. {In Table \ref{tab:combined_inov_stats}, columns 1 $-$ 4 give the SDSS designation of the target source, the observation date, the total number of data points (N) in each DLC and the total monitoring duration (T), respectively. Column 5 presents the computed values of $F_{1}^{\eta}$ and $F_{2}^{\eta}$ for the target's DLCs relative to the two selected comparison stars (indicated in parentheses). The critical values of $F $ ($= F_{c}^{\alpha}$), for $\alpha = $ 0.05 (95\% confidence) and 0.01 (99\% confidence) for INOV detection, are listed in columns 6 and 7, respectively. Column 8 gives the variability status of the two DLCs (relative to the two comparison stars), leading to the overall adopted INOV classification {for $F_{\eta}$ test}, given in column 9. Column 10 specifies the computed F-statistics for the $F_{enh}$ test and column 11 gives the critical value of $F $ ($= F_{c}^{\alpha}$) at the $\alpha = $ 0.01 (99\% confidence) level. Column 12 provides the variability status based on the $F_{enh}$ test. Column 13 details the photometric noise parameter (PNP) and column 14 provides the INOV variability amplitude ($\overline\psi$), {calculated by taking average of the $\psi$ values obtained from the DLCs of the target AGN, for which INOV was detected (type `V'); in the present study this happened only by application of the $F_{enh}$ test. }} Table \ref{tab2:3c} contains additional INOV results for our sample, based on 16 monitoring sessions {for which INOV results} are available in the literature and are directly quoted here. {These 16 sessions include the intranight photometric observations carried out using the 1.04-metre ST telescope and the 1.3-metre JCBT telescope (see \citealp{Goyal2012AA...544A..37G}; \citealp{Pandian2022JApA...43...48P}). {In Table \ref{tab2:3c}, column 1 gives the source name, column 2 the observation date and column 3 mentions the telescope used. Columns 4 and 5 list the total number of data points (N) and the total monitoring duration (T) of the session, respectively. Column 6 specifies the overall INOV classification assigned to each session. Column 7 gives the mean INOV variability amplitude ($\overline\psi$).}}

From Fig. \ref{fig1}, it may be seen that the PSF generally remained steady, i.e. showed no obvious systematic trend, throughout the monitoring sessions. This check is certainly important for avoiding claims of INOV that are spurious, particularly in the case of relatively nearby AGN ($z$ < 0.5, see \citealp{Cellone2000AJ....119.1534C}, also \citealp{Olgu2020MNRAS.492.1450O}), and even for more distant target AGN which happen to have non-negligibly bright optical object(s) projected within a few arcseconds. Here, it should be underlined that the requirement of a steady PSF throughout the session has been duly verified for the supplementary data on INOV results for the 3 sessions targeting the LPSQ 3C 273 ($z$ = 0.158), all using the 1.3-metre JCBT, which we have adopted from the literature
(see above, and Tables \ref{tab:combined_inov_stats} \& \ref{tab2:3c}).\par

A key result emerging from the present analysis is the very low duty cycle (DC) of INOV found for LPSQs. Even though INOV amplitudes down to $2-3\% $ level could have been detected in most of our 39 sessions, setting the INOV detection threshold at $\psi = 5\%$ would be a safer premise to ensure a robust INOV statistics and the same threshold is therefore adopted here. The practice of determining INOV duty cycles for different {thresholds} of $\psi$ was introduced in \citet{GopalKrishna2003ApJ...586L..25G}, as a more objective approach for INOV characterization and it is now widely adopted in the INOV literature. For the present sample of 8 LPSQs, none of the 39 sessions exhibited INOV amplitude above 5\% level, based on the $F_{\eta}$ test (Tables \ref{tab:combined_inov_stats} and \ref{tab2:3c}), implying an INOV DC ($\psi$ > 5\%) of practically zero. 
Only one session exhibited an INOV amplitude approaching the threshold of $\psi = 5\%$. This was for the LPSQ J074110.70+311200.23 monitored on 2006/01/21 ($\psi$ = 4.88\%, Table \ref{tab2:3c}). Inclusion of this session would still yield an INOV DC ($\psi$ > 5\%) of only $\sim$ 3\%.

{Furthermore, the application of the enhanced F-test ($F_{enh}$) to the 23 sessions with data available in digital form, revealed
only one session having INOV detection with amplitude $\psi > 5\%$. This was on 2025/01/25 when the LPSQ J084124.36+705342.17 showed a
$\psi$ of 6.12\% (Table \ref{tab:combined_inov_stats}). Although INOV was also detected for J074110.70+311200.23 observed on 2024/01/30, its amplitude of $\psi$ = 3.77\% is well below the 5\% threshold (Table \ref{tab:combined_inov_stats}). Thus, the INOV duty cycle, DC ($\psi > 5\%$) for our sample is estimated to be $\sim$ 7\%, based on the $F_{enh}$ test.}

To put these findings in perspective, we present below our estimates of DC ($\psi$ > 5\%) for the FSRQs showing a {\it high} degree of optical polarization ($P_{opt} > 3\%$), which are usually termed as HPQs or CDHPQs (i.e. core-dominated HPQs, see \citealp{Stalin2004JApA...25....1S}; \citealp{Goyal2013MNRAS.435.1300G}). 

To do this, we consider two studies, namely \citet{Goyal2013MNRAS.435.1300G} and \citet{Pasierb2020MNRAS.492.1295P}, which report the deduced INOV parameters including $\psi$ for two well-defined sets of HPQs which consist of 11 HPQs {(including 9 low-synchrotron-peaked (LSP); 26 sessions)} and 8 HPQs {(including 5 with LSP designation; 12 sessions)}, respectively. These two studies were considered here because not only they provide the value of $\psi$ for individual sessions, but they also use the same definition of $\psi$ and employ the same $F_{\eta}$ test for INOV detection as {primarily} adopted in this study. Using the R/I-bands information on INOV status provided in those two publications, we have computed DC ($\psi$ > 5\%) for HPQs, using Eq. (\ref{eq:dc}) (Section \ref{stat}), and the values are found to be 27\% and 33.0\%, respectively, considering only the confirmed (and {disregarding} the `probable') cases of INOV detection. It is important to {reiterate} that the DC estimation for the HPQs refers to the LSP subset, defined by $log (\nu_{p}/Hz) < 14$ \citep{Abdo2010ApJ...716...30A}. This is because, {as seen from Table \ref{tab:Table1}, 5 of the 8 sources are specifically classified as belonging to the LSP/LBL type, and the remaining 3 sources are FSRQs, which again are known to overwhelmingly belong to the LSP class (e.g. \citealp{Abdo2010ApJ...716...30A}), also supported by their non-detection at TeV energies (c/f TeVCat\footnote{\url{http://tevcat.org/}}).}  
Thus, the DC estimates for HPQs derived from the above-mentioned two studies are effectively representative of the LSP population of flat-spectrum quasars, allowing a meaningful and valid comparison with our LPSQ sample.

\par

To sum up, a much lower DC ($\psi > 5\%)$ value {(by a factor of 3 $-$ 10)}, is found here for a representative sample of LPSQs (i.e., LPQs exhibiting large superluminal motion in the nuclear jet), as compared to the HPQ subset of FSRQs {(which too have a low-synchrotron-peaked spectrum)}. This shows that a large Doppler boost expected from the observed highly superluminal knots in the nuclear jets of LPSQs does not adequately compensate for the weakness of their intrinsic INOV. This appears to counter to the notion that INOV is primarily linked to the jet's superluminal motion, {as reported in
a recent study, albeit focussed on high-synchrotron-peaked (HSP) blazars (\citealp{Vibhore2023MNRAS.524L..66N}). This distinction may be significant, considering that the two spectral classes are known to differ in optical polarization properties (\citealp{Blinov2016MNRAS.462.1775B}; \citealp{Grigoriou2026A&A...710L..33G}).}

Thus, in summary, the evidence presented here demonstrates that the dependence of INOV on optical polarization ($P_{opt}$) is much stronger, placing the finding by \citet{Goyal2012AA...544A..37G} on a sound footing. The present study has explicitly shown, for the first time, that despite the benefit of a strong Doppler boost associated with the observed large superluminal motion in the nuclear jet, INOV remains subdued if the blazar's optical emission is weakly polarized. Thus, our main conclusion is that at least for powerful jetted AGNs {of low-synchrotron-peaked type}, optical polarization, and not the apparent superluminal speed, is the primary diagnostic of INOV. According to the currently popular paradigm of blazar jets, a high degree of polarization is associated with strong shock(s) forming in the relativistic jet's synchrotron plasma (e.g. \citealp{Blandford1979ApJ...232...34B}; \citealp{Marscher1985ApJ...298..114M}; \citealp{Hughes1991ApJ...374...57H}; \citealp{Marscher1991vagn.conf..153M}; \citealp{Laing1996ASPC..100..241L}). The wake of the shock propagating along the jet is identified as the zone of turbulence excited (or enhanced) by the passing shock, and is the site where INOV is supposed to originate (e.g. \citealp{Marscher1991vagn.conf..153M}; \citealp{Romero2000A&A...360L..47R}; \citealp{Marscher2008Natur.452..966M}; \citealp{Goyal2012AA...544A..37G}; \citealp{Marscher2014ApJ...780...87M}; \citealp{Calafut2015JApA...36..255C}; \citealp{Pollack2016ApJ...820...12P}). The INOV generated inside the turbulent wake appears Doppler boosted, thanks to the jet's bulk relativistic motion towards the observer. Thus, powerful shocks forming inside the nuclear jet seem to play a vital role in boosting both the degree of optical polarization and the turbulence level, the latter inside the jet plasma trailing the shock, where INOV is seeded. Alternative scenarios for the origin of INOV via fluctuating emission pockets widespread across the jet flow, e.g. via distributed magnetic reconnections or magneto-hydrodynamic (MHD) instabilities causing inhomogeneous current sheets (e.g. \citealp{Liang2025A&A...703A.226L} and references therein), which may not be contingent upon powerful shock formation in the jet, seem less compelling in the light of the new observational clue reported here.

\section{Conclusions}
We have reported a sensitive search for intranight optical variability (INOV) in a well-defined, representative sample of 8 flat-spectrum radio quasars (FSRQs) marked by a low degree of optical polarization ($P_{opt}$ < 3\%) but a large superluminal motion
(c$\beta_{app} = 5c - 31c$, median = $10c$). These low-polarization superluminal quasars, called here `LPSQs', were monitored by us in 20 
sessions of a minimum 3-hour duration, using a 1.3-metre telescope. The INOV statistics for this sample could be augmented by including published 
results for some of these blazars, for another 19 intranight monitoring sessions of similarly high-quality, again using 1-metre class telescopes. {The duty cycle (DC) of strong INOV (amplitude $\psi > 5\%$) is found to be zero, according to the normally used $F_{\eta}$ test, and $\sim 7\%$ using the `enhanced F-test' which is a more powerful INOV detector.}
This {small DC} stands in stark contrast to the INOV DC ($\psi$ > 5\%) of $\sim 30\%$ found for the subset of FSRQs showing {\it high} optical polarization, {using the normal $F_{\eta}$ test}. This finding has provided the first clear evidence that optical polarization is the primary diagnostic of INOV, while relativistic beaming plays a supportive role, at least for {powerful jetted AGN displaying a low-synchrotron-peaked spectrum}.
This result is compatible with the `shock-in-jet' model proposed for blazar jets, according to which the passage of a shock along 
the relativistic jet produces strongly polarized synchrotron emission from the vicinity of the shock front generated in the jet-flow, as well 
as a highly fluctuating radiation emanating from the trailing wake of the shock, where the synchrotron plasma is left in a turbulent state. Both these 
components of synchrotron emission appear Doppler boosted due to the jet's bulk relativistic motion towards the observer. Alternative possibilities 
for the origin of optical micro-variability, that invoke a highly inhomogeneous magnetic turbulence reconnection all across the relativistic jet flow 
(see, e.g. \citealp{Liang2025A&A...703A.226L} and references therein), or the magnetic reconnection mini-jet model (e.g. \citealp{Giannios2013MNRAS.431..355G}; \citealp{Sironi2014ApJ...783L..21S}), while also elegant and physically well-motivated, could however become compatible with the present observational clue only provided it can be argued that these widespread dynamic processes occurring in the jet's plasma are inevitably accompanied by a boost in the overall degree of polarization.

\section*{Acknowledgments}
{We thank the reviewer for the helpful comments on the original manuscript.} We are indebted to our late colleague, Mr. Ashutosh Tomar, for his valuable contribution to the observational work presented in this study. This paper is dedicated to his memory. We would also like to thank Prof. M. Gopinathan for making available the archival (raw) data on 3C 273, used in the present study. KC acknowledges the support of Anusandhan National Research Foundation (ANRF), formerly Science and Engineering Research Board (SERB)-Department of Science and Technology (DST), New Delhi, for funding under the National Post-Doctoral Fellowship Scheme through grant no. PDF/2023/004071. G-K thanks the Indian National Science Academy for a Senior Scientist position during which the early part of this work was done. VN is supported by Beijing Natural Science Foundation (Grant No. IS25004). The assistance from the scientific and technical staff of ARIES DFOT is thankfully acknowledged. The local hospitality extended to KC by IUCAA, Pune, during the preparation of this manuscript, and IUCAA associateship to HC are also thankfully acknowledged.

\section*{Data availability}
The data used in this study will be shared at a reasonable request by the corresponding author.


\bibliography{references}



\section*{Appendix}

\begin{figure*}
    \centering
    \begin{minipage}[t]{0.49\textwidth}
        \centering
        \includegraphics[width=\textwidth]{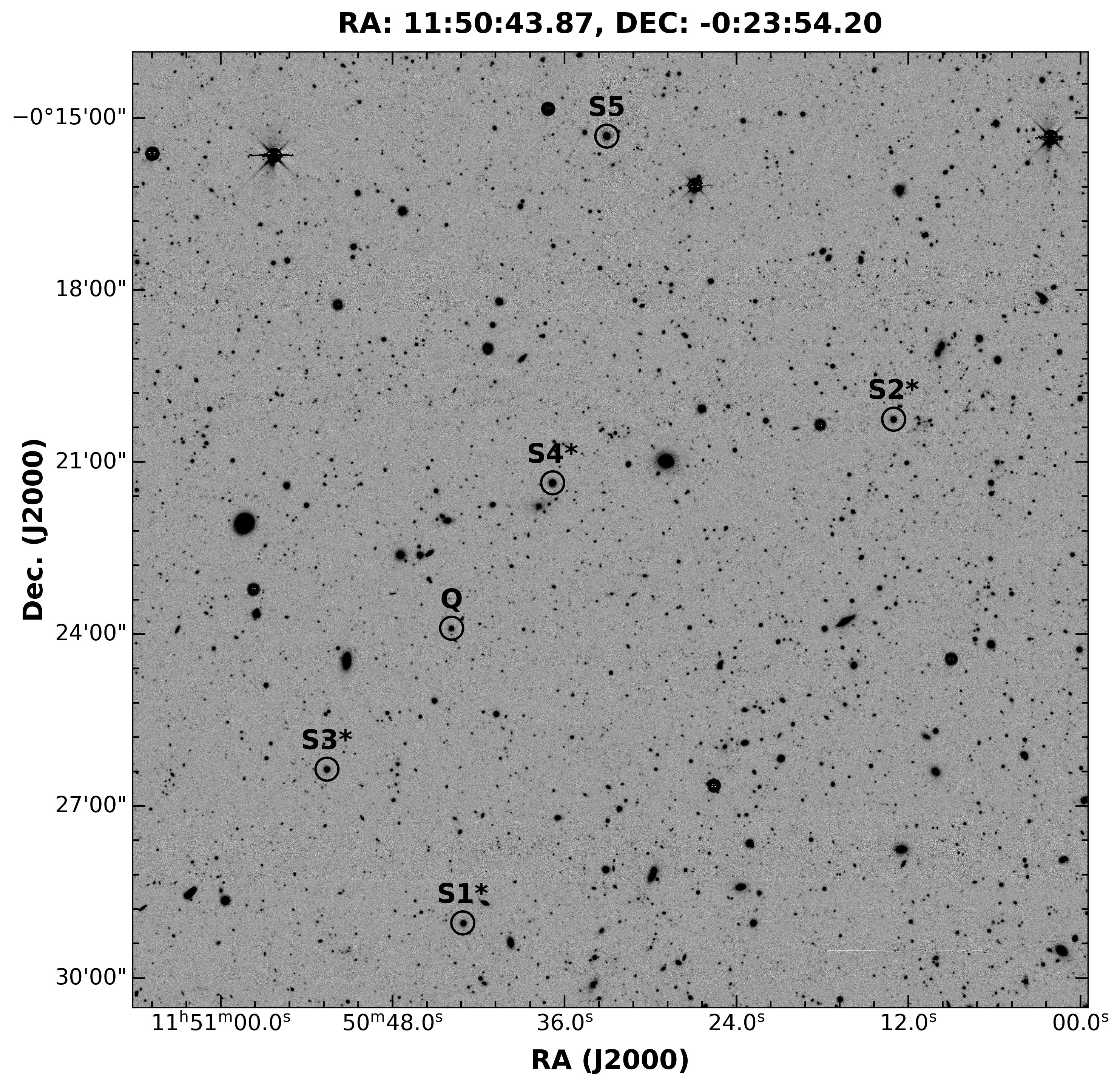}
        \caption{Same as Fig. \ref{fig:J0741_finding}, but for J115043.87$-$002354.20.}
        \label{fig:J1229_finding}
    \end{minipage}
    \hfill
    \begin{minipage}[t]{0.49\textwidth}
        \centering
        \includegraphics[width=\textwidth]{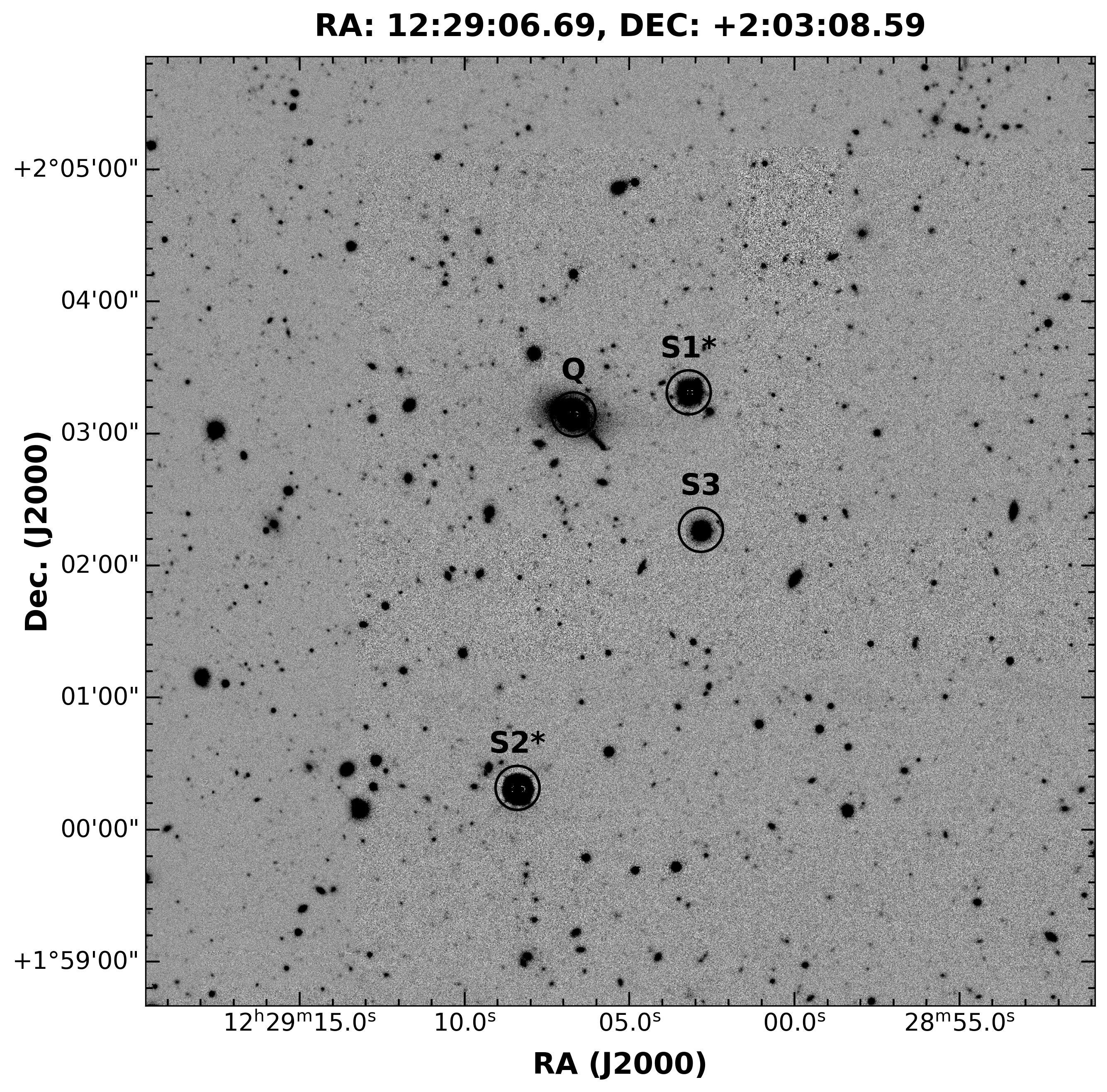}
        \caption{Same as Fig. \ref{fig:J0741_finding}, but for J122906.69$+$020308.59.}
        \label{fig:J1150_finding}
    \end{minipage}
\end{figure*}
   
\begin{figure*}
    \centering
    \begin{minipage}[t]{0.49\textwidth}
        \centering
        \includegraphics[width=\textwidth]{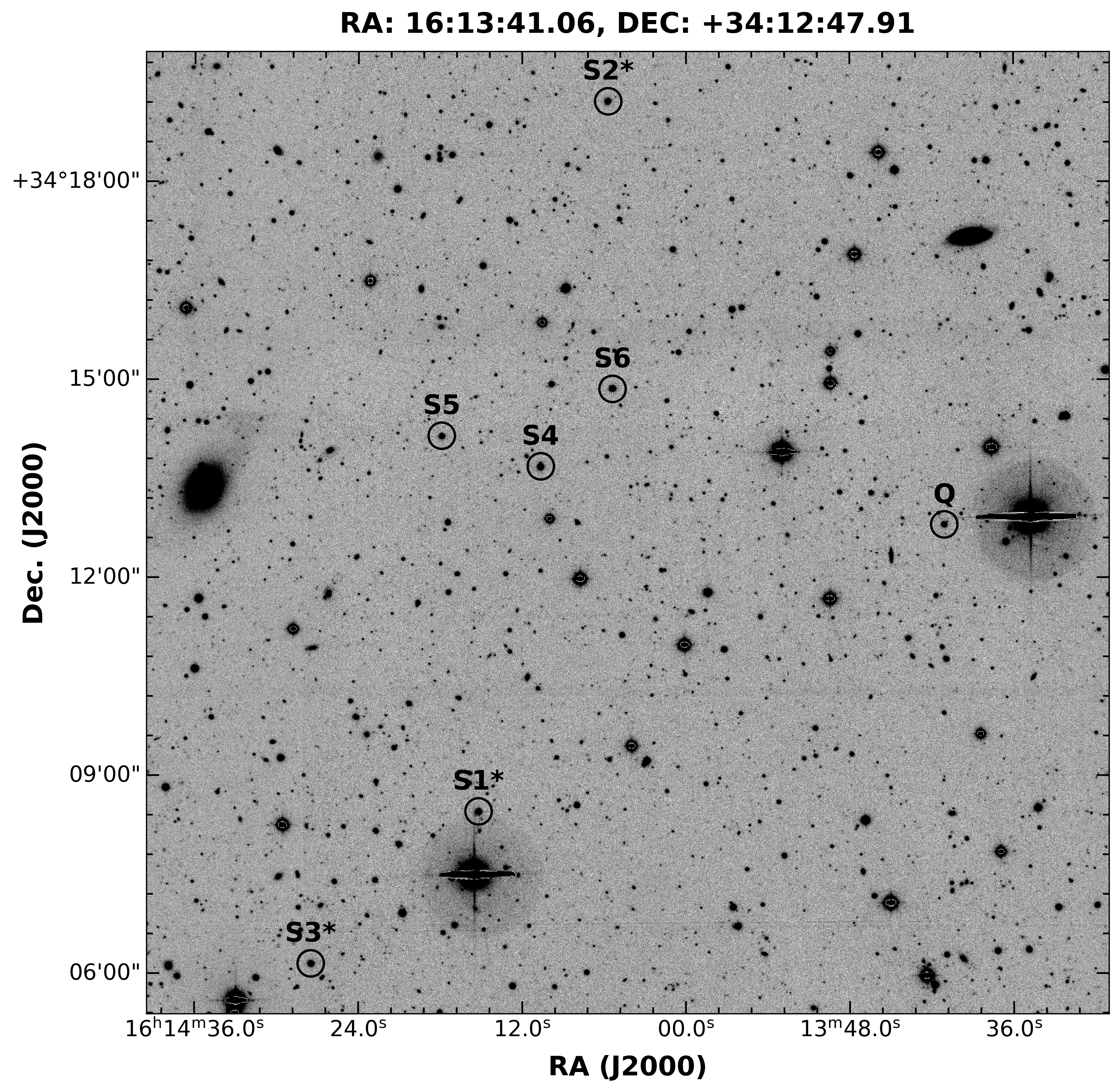}
        \caption{Same as Fig. \ref{fig:J0741_finding}, but for J161341.06$+$341247.91.}
        \label{fig:J1229_finding}
    \end{minipage}
    \hfill
    \begin{minipage}[t]{0.49\textwidth}
        \centering
        \includegraphics[width=\textwidth]{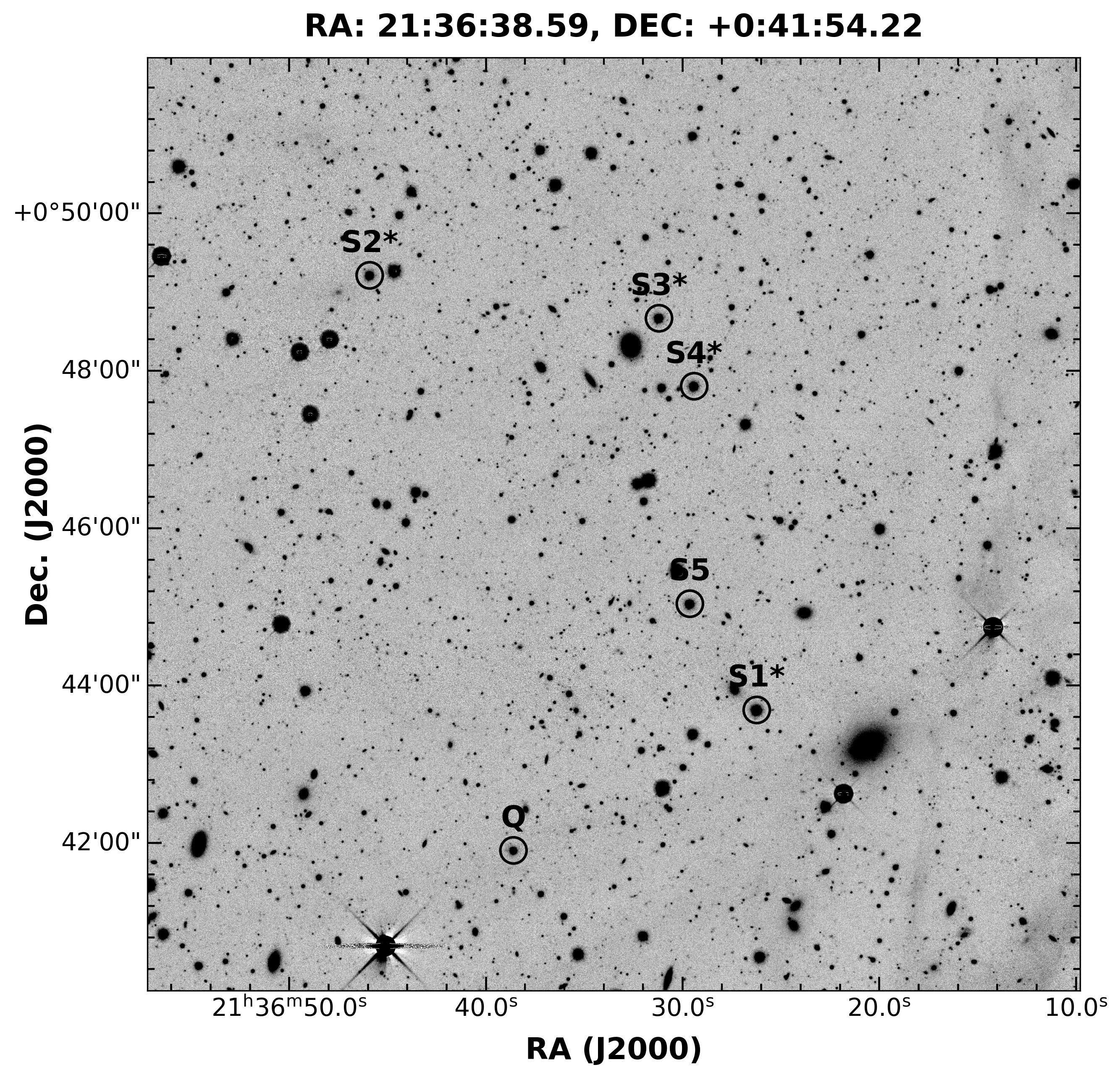}
        \caption{Same as Fig. \ref{fig:J0741_finding}, but for J213638.59$+$004154.21.}
        \label{fig:J1150_finding}
    \end{minipage}
\end{figure*}

\begin{figure*}
        \centering
        \vspace{0pt}
        \includegraphics[width=0.49\textwidth]{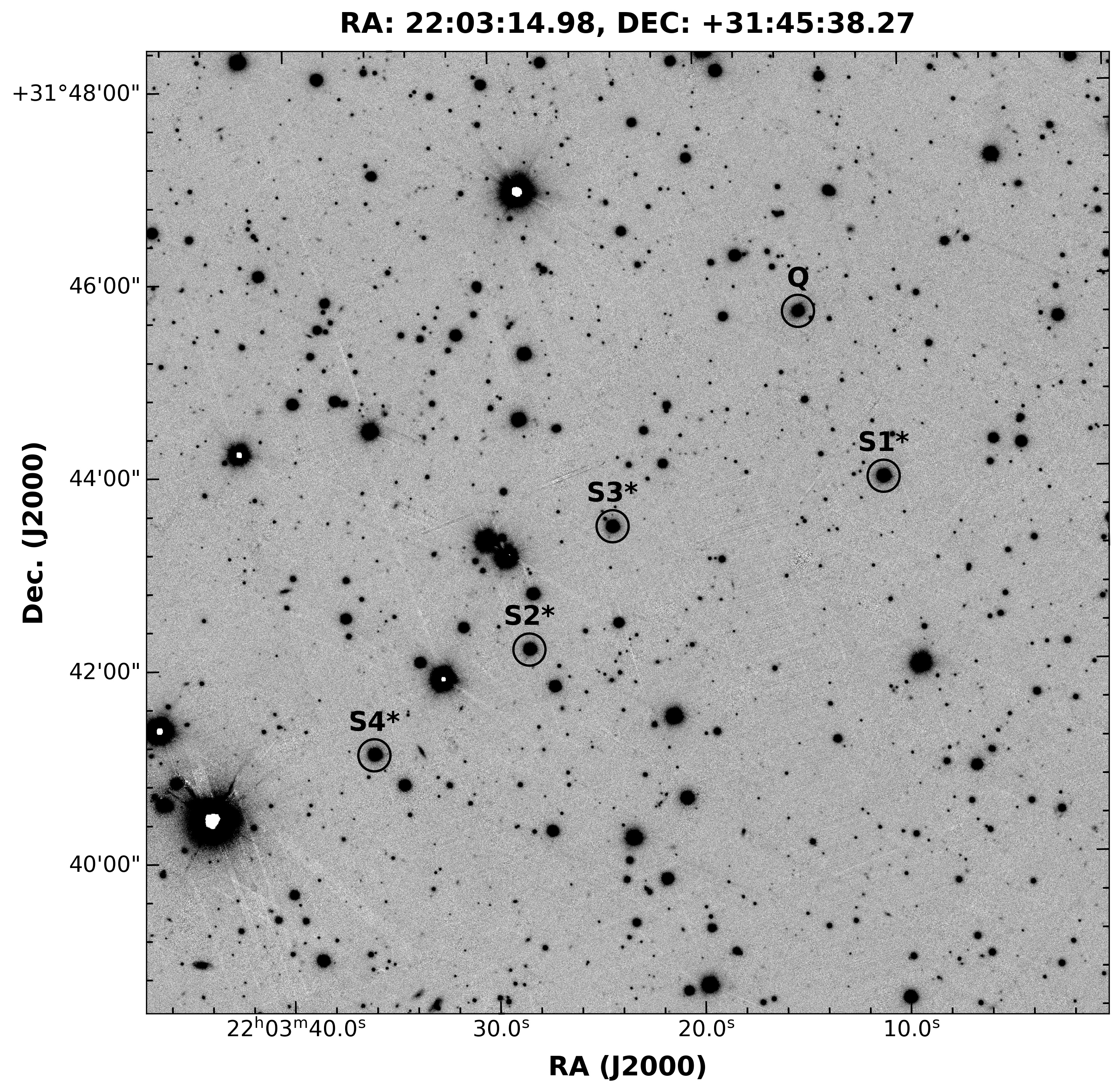}
        \caption{Same as Fig. \ref{fig:J0741_finding}, except for the Pan-STARRS1 (PS1) finding chart for J220314.98$+$314538.27.}
        \label{fig:J22_finding}
        \end{figure*}

\label{lastpage}

\end{document}